\documentclass[aps,pra,epsf,superscriptaddress,amsmath,amssymb,amsfonts,twocolumn,showpacs,nofootinbib]{revtex4-1}
\usepackage{amsmath}
\usepackage{subfigure}
\usepackage{graphicx}
\usepackage{cancel}
\usepackage{booktabs}
\usepackage{dcolumn}
\usepackage{bm}
\usepackage[normalem]{ulem}
\usepackage{amssymb}
\usepackage{soul}
\usepackage{xcolor}
\usepackage{xparse}
\usepackage{color}
\usepackage{enumitem}
\usepackage{enumitem}
\usepackage{pgfplots}
\pgfplotsset{compat=1.9}
\usepackage{tikz}
\usepackage{hyperref}

\usepackage{enumitem}

\usepackage[normalem]{ulem}

\newcommand\red[1]{\textcolor{red}{#1}}

\begin{document}

\preprint{APS/123-QED}


\title{Dispersive shock waves in a one-dimensional droplet-bearing environment}


\author{Sathyanarayanan Chandramouli}
\email{sathyanaraya@umass.edu}
\affiliation{Department of Mathematics and Statistics, University of Massachusetts Amherst, Amherst, MA 01003-4515, USA}

\author{S. I. Mistakidis}
\affiliation{Department of Physics, Missouri University of Science and Technology, Rolla, MO 65409, USA}

\author{G. C. Katsimiga}
\affiliation{Department of Physics, Missouri University of Science and Technology, Rolla, MO 65409, USA}%

\author{P. G. Kevrekidis}
\affiliation{Department of Mathematics and Statistics, University of Massachusetts Amherst, Amherst, MA 01003-4515, USA}%

\date{\today}

\begin{abstract}

We demonstrate the controllable generation of distinct types of dispersive shock-waves emerging in a quantum droplet bearing environment with the aid of step-like initial conditions. Dispersive regularization of the ensuing hydrodynamic singularities occurs due to the competition between mean-field repulsion and attractive quantum fluctuations. This interplay delineates the dominance of defocusing (hyperbolic) and  focusing (elliptic) hydrodynamic phenomena respectively being designated by real and imaginary speed of sound. 
Specifically, the symmetries of the extended Gross-Pitaevskii model lead to a three-parameter family,  
encompassing two densities and a relative velocity, 
of the underlying  Riemann problem utilized herein. 
Surprisingly, dispersive shock waves persist across the hyperbolic-to-elliptic threshold, while a plethora of additional wave patterns arise, such as rarefaction waves, traveling dispersive shock waves, (anti)kinks and droplet wavetrains. The classification and characterization of these features is achieved by deploying Whitham modulation theory. 
Our results pave the way for unveiling a multitude of unexplored coherently propagating waveforms in such attractively interacting mixtures and should be detectable by current experiments.

\end{abstract}

\maketitle


\section{\label{sec:introduction}INTRODUCTION} 

Dispersive hydrodynamics (DH) deals with multiscale wave phenomena in fluid media featuring suppressed bulk dissipation. 
In this context, dispersive shock waves (DSW) are ubiquitous coherent excitations emerging due to the competition between self-steepening nonlinear effects and wave dispersion. 
Their implications extend across various disciplines ranging from surface (and interfacial) waves~\cite{trillo2016observation,maiden2016observation,el2012transformation} and water waves~\cite{ablowitz2011nonlinear,fibich2015nonlinear} in classical settings, to light flow in nonlinear media~\cite{wan2007dispersive,bendahmane2022piston,fatome2014observation}, nonlinear dynamics in cold atomic platforms~ \cite{hoefer2006dispersive,chang2008formation,hoefer2009matter,kevrekidis2008emergent}, and in the materials science
theme of (discrete) granular crystals~\cite{molinari,yasuda2017emergence,chong2018coherent} and related Fermi-Pasta-Ulam-Tsingou lattices~\cite{talcohen}. 

\begin{figure*}
\includegraphics[width=\textwidth]{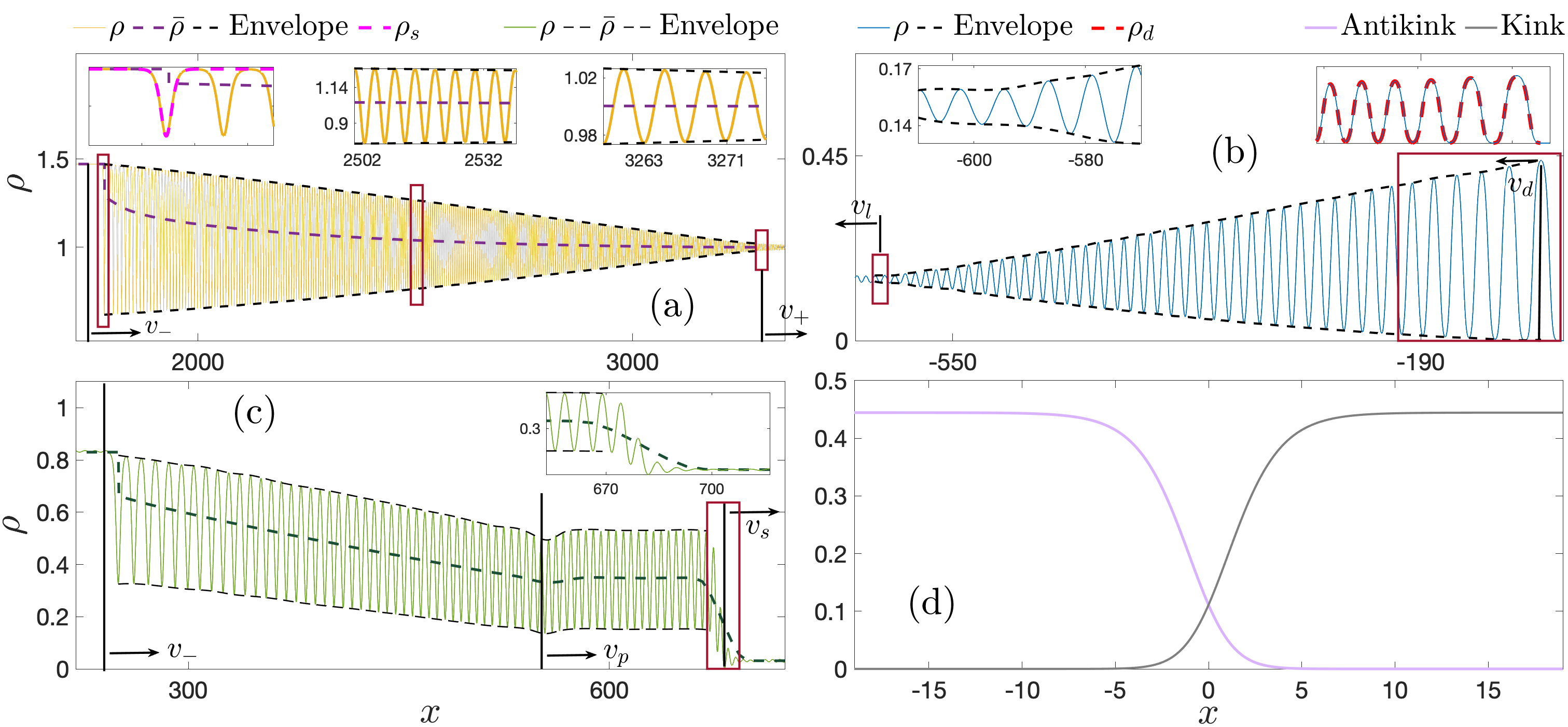}
\caption{{Distinct coherent shock-wave  structures manifesting across the hyperbolic threshold, $\rho_0^{(j)}=0.25$ ($j=1,2$), see Eq.~(\ref{Riemann-step-GP}).  
(a) DSW nucleation as captured by the eGPE dynamics subjected to a dam break problem above the hyperbolic threshold {[($\rho_0^{(1)},\rho_0^{(2)})=(2,1)$]}. Its multiscale structure can be decomposed into three ``zones" indicated by red rectangles. From left to right the insets depict the solitonic edge with speed $v_-$ and fitted by an exact dark-soliton ($\rho_s$) of the eGPE {(see main text)}, the quasi-periodic interior and the linear edge having speed $v_+$. The slowly varying wave mean $\bar \rho$ and envelopes (see legend) reflect the long length scale variation of the entire waveform. (b) Below the hyperbolic threshold, droplet DSWs emerge {(here [($\rho_0^{(1)},\rho_0^{(2)})=(0.15,0)$])}, whose rank-ordered nature is evident by its nearly monotonic envelopes (see legend). 
Rectangles mark the droplet (right inset) and linear edge (left inset) of the pattern with speeds $v_d$ and $v_l$ respectively. In the right inset a sequence of exact 1D droplet solutions, $\rho_d$, are fitted to the first six density oscillations.  
(c) Composite waveforms occurring across the hyperbolic threshold {($\rho_0^{(1)}=2>0.25$ and $\rho_0^{(2)}=0.03125<0.25$)}. They consist of a partial-DSW with edge speeds $v_-$ and $v_p$ situated between the two leftmost vertical lines and a traveling entity with $v_s$. 
The rectangle represents the link of the periodic train to the hydrodynamic background, a magnification of which is provided in the inset. 
The wave envelopes and $\bar \rho$ are also marked. (d) The antikink and kink correspond to a fourth type of shock structure that occurs across the hyperbolic threshold with {$(\rho_0^{(1)},\rho_0^{(2)})=(4/9,0)$ and $(\rho_0^{(1)},\rho_0^{(2)})=(0,4/9)$ respectively.}}}\label{fig:anatomy}
\end{figure*}

The anatomy of the DSW is that of a continuously expanding waveform. The latter refers to a modulated periodic wave encapsulated by a monotonically varying wave envelope, 
extending from a nonlinear (solitonic) edge to
to a linear dispersive wave tail 
(linear edge)~\cite{EL201611}, 
see also Fig.~\ref{fig:anatomy}(a). 
In essence, DSWs are generated either as a consequence of the dispersive regularization (Riemann problem) of wave breaking (gradient catastrophe)~\cite{bendahmane2022piston} or when the ``flow" speed is close to the speed-of-sound intrinsic to the medium \cite{grimshaw1986resonant,hakim1997nonlinear}, for further details see also review~\cite{EL201611}. 
A standard framework to unravel the behavior of DSWs is the celebrated  
Whitham modulation theory~\cite{whitham1965non}. 
In essence, this method is a multiscale approach and refers to averaging over the fast oscillatory scale. 
This process leads to a set of spatiotemporal modulation equations incorporating the slow variation of the parameters governing the multiscale periodic wave such as the DSW. 
The number of these parameters has to match the number of conservation laws, so as to obtain a meaningful dimensional
reduction through such a scheme.

Cold atom simulators have been proven to be fertile
playgrounds for investigating coherent nonlinear structures, see the reviews of~\cite{bloch2008many,polkovnikov2011colloquium,kevrekidis2015defocusing}, owing to their high degree of parameter tunability and isolation. 
Here, intense research activity has been focused on DSW generation~\cite{Damski1, Damski2,hoefer2006dispersive,proukakis2006quasicondensate} spearheaded by their experimental observation in Bose-Einstein condensates~\cite{dutton2001observation} and later on in Fermi gases~\cite{joseph2011observation}. 
Properties of DSWs have also been discussed in long-range interacting setups such as strongly interacting Rydberg settings~\cite{hang2023accessing}. 
In the same spirit, more recently, DSW nucleation has been reported within an extended Gross-Pitaevskii (eGPE) model~\cite{petrov2015quantum} featuring contact interatomic interactions as a by-product of kink-antikink interactions~\cite{katsimiga2023interactions}. 
This model, besides the standard cubic nonlinearity incorporates the effect of the first order Lee-Huang-Yang (LHY) quantum correction~\cite{lee1957eigenvalues} which,
in the effectively one-dimensional setting, introduces 
(an attractive) quadratic nonlinear term. 

Within this eGPE model another state of matter forms the so-called quantum bright droplets which have been experimentally observed in both homonuclear~\cite{cheiney2018bright,semeghini2018self,cabrera2018quantum} and heteronuclear~\cite{d2019observation,burchianti2020dual} bosonic mixtures (at least in higher dimensional
settings). 
They refer to many-body self bound states of matter~\cite{luo2021new,mistakidis2023few,khan2022quantum} existing due to the competition between mean-field (MF) repulsion and quantum fluctuation attraction (represented
by the quadratic nonlinearity in one-dimension (1D)~\cite{petrov2016ultradilute,mistakidis2023few}, a situation that is reversed in three-dimensions~\cite{petrov2015quantum}. 
Yet, additional coherent structures can be hosted in this setup such as bubbles~\cite{katsimiga2023interactions,katsimiga2023solitary,edmonds2023dark} (also known as dark droplets), kinks~\cite{kartashov2022spinor,katsimiga2023interactions}, dark solitons~\cite{katsimiga2023solitary,saqlain2023dragging,gangwar2022dynamics} and vortices in two-dimensions~\cite{kartashov2022spinor,tengstrand2019rotating,li2018two}. 
Notice that droplets (bubbles) are known to be stable~\cite{tylutki2020collective,du2023ground} (unstable~\cite{katsimiga2023interactions}) structures, solitary waves feature  parametric windows of instability~\cite{katsimiga2023solitary}, while plane waves are modulationally unstable leading to droplet mergers~\cite{mithun2020modulational,mithun2021statistical}.
Interestingly, it was demonstrated~\cite{katsimiga2023interactions} that DSWs dynamically emerge as a by-product of the interaction between some of the aforementioned coherent entities including droplets and kinks, kink with antikinks but also through the destabilization of bubbles. 
However, the origin and properties of DSWs in droplet related systems characterized by the interplay of the qubic-quadratic interactions consist open questions. 
Such an interaction interplay for studying DSWs has been exploited in a different context of the cubic-quintic nonlinear Schr\"odinger (NLS) equation~\cite{crosta2012whitham}.

It is the main purpose of our work to systematically study and control the dynamical generation of DSWs in such a 1D droplet supporting environment modeled by an eGPE. 
For this reason, a piecewise constant step initial condition featuring zero velocity, referred to as dispersive ``dam break problem", arising also in shallow water theory~\cite{whitham2011linear,leveque2002finite}, is deployed. Interestingly, it is found that the interplay of the involved interaction terms leads to mean-field and LHY dominated dynamical response regimes characterized primarily by real and imaginary speed of sound. The latter is analytically extracted through the hydrodynamic reduction of the model under consideration.  
In the former regime, the spontaneous generation of robustly propagating rarefaction waves and DSWs is observed, manifesting the repulsive mean-field  dynamics~\cite{hoefer2014shock,el1995decay}. 
Excellent agreement between the predictions of the eGPE model and the analytical ones stemming from the so-called Whitham-El closure method~\cite{hoefer2014shock,el2005resolution} is demonstrated. 
This includes the characterization of the rarefaction wave profiles and macroscopic dynamical properties of the emitted wave patterns such as the velocity and density of the intermediate background, and DSW edge speeds. 

Remarkably, across the threshold between the mean-field and LHY dominated regimes, we find different flavors of ``shock waves". The latter include, among others, Whitham shocks with (anti-)kinks being  identified for the first time as belonging to this larger family. Moreover, there exist DSWs that are seen to persist across the obtained 
(real-to-imaginary speed of sound) threshold which we term as DSW remnants. Additionally, deep within the LHY dominated response regime, we observe robustly propagating droplet DSWs, which are reminiscent of their bright solitonic  counterparts arising in optical settings~\cite{biondini2018universal,el2016dam,wan2010diffraction} and attractive superfluids \cite{kh2005dynamics}. 
The characteristic time of initiation of the droplet DSWs appears to be related to modulation instability (MI) of plane waves. It turns out that all the aforementioned DSW structures, even though subjected to MI in one of the asymptotic plane wave backgrounds, are surprisingly long-lived. The quantum droplet bearing system thus provides a versatile environment to access distinct regimes in which different dispersive hydrodynamic features nucleate spontaneously. 

This work unfolds as follows. 
In Section~\ref{setup} we introduce the reduced single-component eGPE model supporting droplet solutions and describe the initial condition (Riemann problem) allowing the  dynamical emergence of DSWs.  
Section~\ref{hydrodynamics} elaborates on the hydrodynamic formulation of the eGPE, the conditions under which MI appears in this system and its underlying conservation laws. 
In Section~\ref{sec:results} we discuss the spontaneous generation of DSWs and their characteristics in the droplet environment. 
Whitham modulation theory is deployed to interpret the properties of DSWs obtained from the simulations of the eGPE model.  
We conclude and offer possible future research extensions of our findings in Section~\ref{conclusions}. In Appendix~\ref{appendix:Whitham_eq} we lay out some analytical formulations related to the Whitham-El closure method \cite{el2005resolution} and also briefly touch upon a few instances of general Riemann problems with non-zero hydrodynamic velocity.

\section{Setup $\&$ initial condition}\label{setup}

The setup under consideration consists of a 1D homonuclear bosonic 
mixture. The atoms reside in two different hyperfine states 
of, e.g., $^{39}$K as per the corresponding 3D experiment of Ref.~\cite{semeghini2018self} in free space. 
The 1D geometry is achieved by utilizing tightly confined transversal directions as compared to the elongated un-confined $x$-direction. 
For simplicity, both states share the same atom number ($N_1=N_2\equiv N$). They feature equal intracomponent repulsion,  $g_{11}=g_{22}\equiv g>0$, whereas intercomponent attraction, $g_{12}<0$, gives access to the droplet environment for {$\delta g=g_{12}+g>0$}.  
These assumptions render the participating components equivalent and the mixture can be described by a reduced single-component eGPE~\cite{petrov2015quantum,petrov2016ultradilute} which incorporates the first order LHY quantum correction. 
The relevant reduced dimensionless model reads  
\begin{equation}
\label{GP-eqn}
i\psi_t+\frac{\psi_{xx}}{2}-|\psi|^2\psi+|\psi|\psi=0,
\end{equation}
where $\psi$ is the 1D wave function. 
Importantly, the quadratic nonlinearity encompasses the 1D attractive nature of the LHY contribution, whereas the cubic term accounts for the standard mean-field repulsion. 
Here, the energy of the system is expressed in terms of $\hbar ^{2}/(m\xi ^{2})$, with $\xi =\pi \hbar ^{2}\sqrt{|\delta g|}/(mg\sqrt{2g})$ denoting the healing length. The atom mass is $m$ and $\hbar$ is the reduced Planck constant. Also, time, length, and wave function are in units of  $\hbar/\left(m\xi ^{2}\right) $, $\xi $ and $(2\sqrt{g})^{3/2}/(\pi \xi (2|\delta g|)^{3/4})$  respectively. 
Typical evolution times considered herein are of the order of $t\sim 10^3$ translating to $\sim 800 $ms e.g. for a transverse confinement $\omega_{\perp}\approx 200$Hz used in the experiment of Ref.~\cite{semeghini2018self}.    

This 1D eGPE model admits a plethora of coherent structures
such as droplets~\cite{petrov2015quantum,tylutki2020collective}, bubbles~\cite{katsimiga2023interactions}, 
single- and multiple-dark soliton states~\cite{edmonds2023dark,katsimiga2023solitary,katsimiga2023interactions}
but also stationary periodic waves of the Jacobi-elliptic type, traveling periodic waves, and kinks see e.g. Refs~\cite{mithun2020modulational,katsimiga2023solitary,katsimiga2023interactions}. 
Additionally, this model has been used to study interactions among several of the aforementioned entities~\cite{katsimiga2023interactions}, but it was also utilized to explicate relevant MI phenomena~\cite{otajonov2022modulational,mithun2020modulational,mithun2021statistical}. The existence of periodic travelling waves and information regarding modulationally unstable parametric regions are of vital importance for the generation and subsequent propagation of DSWs that are the focus of our current investigation.

To unravel the spontaneous nucleation of DSW structures subject to the eGPE [Eq.~(\ref{GP-eqn})] but also understand their characteristics and longevity we utilize a three parameter family of Riemann problems. 
In this context, the Riemann problem refers to the 
evolution dynamics of a piecewise constant waveform composed of two distinct (i.e., $j=1,2$) ``hydrodynamic" backgrounds. The latter have the general form $\psi_0^{(j)}=\sqrt{\rho_0^{(j)}}\exp\left[i(u_0^{(j)} x-\Omega t)\right]$ with $\Omega=\frac{1}{2}(u_0^{(j)})^2+\rho_0^{(j)}-\sqrt{\rho_0^{(j)}}$ and $\rho_0^{(j)}$ ($u_0^{(j)}$) designating the constant amplitude (velocity) of the waveform. We remark that Eq.~\eqref{GP-eqn} possesses translational and Galilean invariance but also reflection~\footnote{Here, $\psi(x-u_0 t,t)e^{iu_0 (x-\frac{u_0}{2})t}$ (Galilean
transformation to a traveling solution with speed $u_0$), and $\psi(-x,t)$ (reflection) satisfy Eq.~\eqref{GP-eqn}.} symmetry. 
However, the loss of scaling invariance is evident due to the presence of two  distinct nonlinearities, namely cubic and quadratic. This necessitates the existence of the second parameter $\rho_0^{(2)}$ in the Riemann initial condition, since it cannot be scaled out. 
It is the above symmetry considerations that lead to the three-parameter family $\{\rho_0^{(1)}$, $\rho_0^{(2)}$, $u_0^{(1)} \}$ of the ensuing Riemann problem. This can be adequately described, setting $u_0^{(2)}=0$ without loss of generality, by the following step-wise initial ansatz 
\begin{align}
\label{Riemann-step-GP}
\psi(x,0)=
\begin{cases}
\sqrt{\rho_0^{(1)}} e^{i u_0^{(1)} x},\;x<0\\
\sqrt{\rho_0^{(2)}}, \; x>0.
\end{cases}
\end{align}
In the remainder, also without loss of generality, we choose $\rho_0^{(1)}>\rho_0^{(2)}$. 
This appropriate initial condition allows us to explore the impact of the three parameter space of Riemann problems of Eq.~\eqref{GP-eqn} on the DSW dynamical generation. Note that the case of $u_0^{(1)}=0$, that we will mainly focus on below, is known as a dispersive ``dam-break" problem~\cite{whitham2011linear,leveque2002finite}. {Experimentally, such an initial condition characterized by density asymmetry around a focal point can be achieved for instance by utilizing phase masks~\cite{pedaci_positioning_2006} or digital micromirror devices~\cite{navon_quantum_2021}. These  allow to imprint arbitrary shapes even in box potentials. Another possible way would be to use a moving repulsive potential barrier (build-up via an optical dipole force of a far detuned laser beam) as it was demonstrated in Ref.~\cite{mossman2018dissipative} for producing DSWs in repulsive condensates. Finally, the initial velocity can achieved, for instance, by using magnetic field gradients to induce a directional flow.}

The spatiotemporal evolution of the eGPE is tracked using a second-order finite differences method for the spatial derivatives and a fourth-order Runge-Kutta~\cite{quarteroni2010numerical} for the dynamics with time discretization $dt = 10^{-4}$. 
To model the Heaviside step-wise initial condition we employ relatively smoother ramps of the $\tanh(ax)$ type with $a \gg 1$.  
To ensure that edge effects are suppressed during the evolution an adequately large spatial domain is employed. The underlying spatial discretization is of the order of $dx=0.01$ enabling high resolution of the generated localized features. 
We have also cross-checked the robustness of the dynamical generation of DSWs in our setting with a pseudo-spectral time-stepping method based on the modified exponential time-differencing method \cite{kassam2005fourth} for representative cases.

\section{Dispersive hydrodynamic formulation }\label{hydrodynamics}

By introducing the Madelung transformation $\psi=\sqrt{\rho(x,t)}\exp(i \int_{x_0}^{x}u dy)$ ~\cite{el2016dispersive}, the eGPE [Eq.~\eqref{GP-eqn}] assumes 
the following DH form 
\begin{subequations}
\begin{align}
\label{Quasilinear-system-DH-1}
    & \rho_t+(\rho u)_x=0,\\
    \label{Quasilinear-system-DH-2}
    & u_t+uu_x+\frac{1}{\rho}{\mathsf{P}_x}=\left(\frac{2\rho\rho_{xx}-\rho_x^2}{8\rho^2}\right)_x.
\end{align}
\end{subequations}
Here, $\rho \equiv |\psi|^2$ represents the droplet  density and $ u={\rm Im}(\psi^{*}\psi_x)/|\psi|^2$ denotes  the corresponding velocity which can also be expressed in terms of the phase $\phi(x,t)$ of the droplet as $u(x,t)=\partial \phi(x,t)/\partial x$. 
The term $\mathsf{P}(\rho)=\left(\rho^2/2-\rho^{3/2}/3\right)$ is referred to as the hydrodynamic pressure defining the speed of sound $c=\sqrt{d\mathsf{P}/d\rho}=\sqrt{\rho-\sqrt{\rho}/2}$. The system of Eqs.~\eqref{Quasilinear-system-DH-1}-\eqref{Quasilinear-system-DH-2} possesses reflection and translation symmetries as well as Galilean invariance (see also the discussion below). 
Finally, the dispersive term on the right-hand side of Eq.~\eqref{Quasilinear-system-DH-2} defines
the so-called ``quantum pressure"~\cite{pethick2008bose} given by $-{(\sqrt{\rho})_{xx}}/{2\sqrt{\rho}}$. 
Note in passing, that in the absence of the quantum pressure term, the aforementioned system of hydrodynamic equations resembles the ones governing the isentropic gas dynamics and shallow water fluid flow~\cite{whitham2011linear}.

In the context of Eqs.~\eqref{Quasilinear-system-DH-1}-\eqref{Quasilinear-system-DH-2}, rescaling of the independent variables $X=\epsilon x$ and $T=\epsilon t$, where $\epsilon$ is a formally small parameter ($\epsilon \ll 1$) reveals that to leading order $\sim\mathcal{O}(\epsilon)$ referring to short times, i.e., $\sim\mathcal{O}(T)$ the quantum pressure term can be neglected. 
Furthermore, the speed of sound can acquire either real or complex values.  
In the former case, the system reduces to a purely hyperbolic one 
describing the behavior of a classical fluid~\cite{leveque2002finite} and can be cast into a Riemann invariant form 
\begin{equation}
\label{Hydrodynamic-reduction}
    \partial_T r_{i}+a_{i}(\vec{r})\partial_X r_i=0,\;i=1,2.
\end{equation}
The Riemann invariants $\Vec{r}=(r_1,r_2)$ are given by $r_{1,2}=(u/2)\mp \int (1/2\rho)c(\rho) d\rho$ [see also Eq.~\eqref{Expression-Riemann} in the Appendix~\ref{appendix:Whitham_eq}],  while the associated characteristic velocities correspond to $a_{1,2}=u\mp c(\rho)$. 
Specifically, $c$ is real (complex)   
for $\rho\geq 1/4$ ($\rho<1/4$) leading to a hyperbolic (elliptic) reduction of Eq.~\eqref{Hydrodynamic-reduction}. Importantly, this hyperbolic (elliptic) behavior is related to dominance of mean-field repulsion (LHY attraction). Indeed, in the absence of the LHY contribution the elliptic regime does not exist.

A fundamental question in the study of DH  systems is the stability of plane waves on which various nonlinear excitations, e.g., solitary waves, propagate. 
The dual hyperbolic-elliptic character of the hydrodynamic reduction [Eq.~\eqref{Hydrodynamic-reduction}] has implications for this stability.
To explore the latter, i.e., study the propagation of infinitesimal disturbances on the plane-wave, we next linearize Eq.~\eqref{GP-eqn} about a plane wave solution $\psi_0(x,t)=\sqrt{\rho_0}\exp(i u_0 x-i\Omega t)$,  extracting in this way the following dispersion relation 
\begin{equation}
\label{Disp-relation-linear-edge}
\omega_0(k,\rho_0,u_0)={u}_0 k\pm k\sqrt{\frac{k^2}{4}+\rho_0-\frac{\sqrt{\rho_0}}{2}}.
\end{equation}
As already highlighted above, in the elliptic regime of the hydrodynamic limit $\rho_0-\sqrt{\rho_0}/2<0$. This leads to a complex dispersion (and thus imaginary $\omega_0$) for sufficiently long wavenumbers $0<k<\sqrt{2\sqrt{\rho_0}-4\rho_0}$ implying MI of
plane waves in the quantum droplet environment as
reported previously, e.g., in Refs.~\cite{mithun2020modulational,mithun2021statistical}. However, in the hyperbolic regime,  plane waves are modulationally stable. 
Such dispersive hydrodynamic systems featuring dual elliptic-hyperbolic regimes have received attention in the context of cubic-quintic NLS equations modeling double shock-wave phenomena in optical and superfluid media~\cite{crosta2012crossover,crosta2012multi,crosta2012double}. 

Moreover, this DH system (and thus the eGPE) exhibits local conservation of i) density 
$\mathbb{P}_1=|\psi|^2$, ii) momentum $\mathbb{P}_2={\rm Im}(\psi^{*}\psi_x)$, and iii) energy  $\mathbb{P}_3=\frac{1}{2}\left(|\psi_x|^2+|\psi|^4\right)-\frac{2}{3}|\psi|^3$. 
The corresponding density, momentum, and energy fluxes are $\mathbb{Q}_1={\rm Im}(\psi^{*}\psi_x)$, $\mathbb{Q}_2= (({\rm Im}(\psi^{*}\psi_x))^2/|\psi|^2)+\mathsf{P}(|\psi|^2)-\frac{|\psi|^2 ({\rm ln}|\psi|^2)_{xx}}{4}$ and $\mathbb{Q}_3=-\frac{1}{2}\left(\psi_x^{*}\psi_t+\psi_x\psi_t^{*}\right)$ respectively\footnote{The total atom number and energy (i.e. $\int_{-\infty}^{\infty} \mathbb{P}_{i} dx$ for $i=1,3$) are conserved throughout the evolution featuring relative errors $\sim 10^{-13}$ and $\sim 10^{-6}$. The integrated momentum should change at a constant rate $\frac{d}{dt}\int_{-\infty}^{\infty} \mathbb{P}_{2} dx=\mathsf{P}(\rho_0^{(1)})-\mathsf{P}(\rho_0^{(2)})$ and this is captured up to a relative error of $\sim 10^{-9}$.} satisfying the continuity equation 
\begin{equation}
\label{mass}
\partial_t \mathbb{P}_i+\partial_x {\mathbb{Q}_i}=0, \;i=1,2,3.
\end{equation}
The aforementioned conservation laws along with the existence of traveling waveforms that the eGPE admits~\cite{katsimiga2023interactions,mithun2020modulational} are the pre-requisites for studying DH features such as the DSW~\cite{EL201611,hoefer2014shock}. 
Recall that the DSW is a modulated periodic wave encompassing a multitude of nonlinear dispersive oscillations and it is commonly studied through the lens of Whitham modulation theory~\cite{whitham2011linear}. 

Due to its multiscale rank-ordered nature, the DSW in convex dispersive hydrodynamic theory~\cite{el2005resolution,el2016dispersive} possesses both microscopic (internal oscillations) and macroscopic characteristics (e.g. distinct speeds and amplitudes associated with its edges). {In what follows, we aim to describe the anatomy of various types of shock waves that arise within the eGPE model {whose characteristics are outlined in Table \ref{tab:notes} and their density profiles are shown in Fig.~\ref{fig:anatomy}.} Concretely, we choose specific spatial domains that solely focus on the individual shock wave nucleation subject to Eq.~\eqref{Riemann-step-GP} and the distinct constituents appearing in each of them. Four out of five in total different types of shock waves that will be reported in what follows, are illustrated in Fig.~\ref{fig:anatomy}. These correspond to the (standard) \textit{DSW} [Fig.~\ref{fig:anatomy}(a)], \textit{droplet DSW} [Fig.~\ref{fig:anatomy}(b)], \textit{traveling DSW} [Fig.~\ref{fig:anatomy}(c)] and the \textit{(anti)kink} [Fig.~\ref{fig:anatomy}(d)], see also Section~\ref{sec:results} for their detailed characterization.} {Finally, yet another waveform identified herein is the so-called $DSW$ remnant. Since it is structurally similar with the standard $DSW$, it is not  shown in Fig.~\ref{fig:anatomy} for brevity, but rather discussed in detail in Section~\ref{sec:results}, see also  Table~\ref{tab:notes}.} 

{A paradigmatic example of an eGPE DSW, whose density profile develops above $\rho > 0.25$ (hyperbolic-to-elliptic threshold) of the hydrodynamic reduction [Eq.~\eqref{Hydrodynamic-reduction}], is depicted in Fig.~\ref{fig:anatomy}(a). As expected, such a waveform emerges in a mean-field (repulsive) dominated regime.}
The macroscopic characteristics describe the ``overall" features of the wavetrain, which include prominently the velocity of its left, $v_{-}$, and its right, $v_{+}$,  bounding edges (see relevant discussion in Sec.~\ref{Mean-field-driven-dynamics}). The difference in edge velocities $w=v_+-v_-$ then describes the rate of spreading of the wavetrain. To capture the rank-ordered nature of the DSW over its slow modulation scale, a monotonic envelope containing the rapid oscillations has been overlaid indicated by black dashed lines. Three different ``zones" of the DSW can be discerned in Fig.~\ref{fig:anatomy}(a) marked by red rectangles. These are the solitonic edge, the modulated periodic
interior, and the linear edge comprised of very small amplitude oscillations {(see also Section~\ref{sec:results}).} 
 
 Particularly, the leftmost inset depicting the first zone provides a magnification of the solitonic edge being compared to an exact eGPE dark soliton computed using a Newton scheme \cite{yang2009newton} (see the magenta dashed line). {To be concrete, the fixed point to an ordinary differential equation upon substituting the traveling wave profile $\Phi(\xi)\exp[i(\theta(\xi)-(\rho_m-\sqrt{\rho_m})t)]$ to the eGPE \eqref{GP-eqn} is identified. Here, $\Phi$, $\theta$ are the amplitude and phase profiles respectively, while $\xi\equiv x-c_s t$ denotes the traveling coordinate. 
 Also, $\rho_m$ is the soliton background, with velocity $u_m$ and $c_s=v_--u_m$ is the soliton velocity in the reference frame of the background (see also below).} {Thereafter, we emphasize that while the subsequent 2-3 oscillations are soliton-like, they are part of the modulated periodic DSW interior. This is rooted in the gradual shift in the nonlinear and self-similar character of a DSW as we traverse from the solitonic to the linear edge.} The middle and right insets present a zoomed view of the periodic oscillations in the DSW interior and linear edge respectively. Further encapsulating the multi-scale nature of this DSW, the corresponding wave mean, $\bar \rho$, varying slowly across the DSW structure, is shown by the purple dashed line. 
 
{Crucially, the eGPE also features other regimes bearing distinct ``shock wave" flavors. }This is tied to the dual elliptic-hyperbolic behavior associated with the hydrodynamic limit of Eqs.~\eqref{Hydrodynamic-reduction}. One such feature is a DSW-like train that emerges from a gradient catastrophe (discontinuity in the initial condition) in the elliptic regime [Fig.~\ref{fig:anatomy}(b)]. This wavetrain mimics the form of a DSW with a rank-ordering of its constituent waves demonstrated by its nearly monotonic envelope (black dashed line). Moreover, there exist two distinct edges. The nonlinear (linear) edge propagates with intrinsic speed $v_d$ ($v_l$) leading to wavetrain spreading. However, several density oscillations within this wavetrain are droplet-like. This can be seen in the right inset of Fig.~\ref{fig:anatomy}(b),  where a {density} fitting of the analytical 1D localized droplet solution (dashed blue line)~\cite{petrov2016ultradilute,tylutki2020collective,katsimiga2023solitary} given by

\begin{equation}
\label{Droplet-solution}
   { \rho_d= \bigg(\frac{3\mu}{1+\sqrt{1+\frac{9\mu}{2}}\cosh(\sqrt{-2 \mu}x)}\bigg)^2,}
\end{equation}
{with $\mu\in (-2/9,0)$ being the chemical potential, is performed. These are quite distinct in character to dark solitons, whose density profile is ${\rm tanh}^2(x)$-like, constituting ``bright'' states on top of
a vanishing amplitude, rather than ``dark'' ones
on top of a finite amplitude. While the relevant distinction might be less straightforward to discern
at the small amplitude limit on one edge of the pattern,
it is definitively discernible at the large amplitude
limit of the opposite end of the pattern.}
Specifically, six such droplets are illustrated showcasing excellent agreement with the wavetrain's tail. This scenario of several modulated density oscillations approaching the droplet limit occurs 
(and evolves) dynamically. This is reminiscent of the saturation of modulated oscillations to bright solitons being observed in the context of attractive interaction (i.e., focusing) NLS models \cite{biondini2018universal}. 
Moreover, the left inset in Fig.~\ref{fig:anatomy}(b) provides a zoom of the relevant linear edge. 
Notice that the generation mechanism of the droplet DSW also differs significantly from the standard mean-field DSW, see the discussion in Sec.~\ref{LHY-driven-evolution}. 
 
A third type of ``shock waves", presented in Fig.~\ref{fig:anatomy}(c), can emerge from the gradient catastrophe across the hyperbolic threshold of $\rho=0.25$. Here, a composite waveform, {distinct from a standard DSW}, comprising a ``partial" DSW, i.e., a DSW that only contains an incomplete range of nonlinear dispersive oscillations marked by vertical black lines, and a peculiar traveling wave feature occurs. This composite entity experiences a slowly varying wave mean $\bar \rho$ across the partial-DSW that saturates close to $\bar \rho\approx 4/9$ before rapidly dropping below ${\bar \rho}\approx 0.1$ (green dashed line). The relevant waveform velocities in this case correspond to that of the soliton edge, $v_-$, the right partial DSW edge, $v_p$, and the traveling wave one, $v_s$. 
Moreover, there exists a fourth member of the ``shock wave" family, shown in Fig.~\ref{fig:anatomy}(d). Indeed, we will showcase in what follows that these well-known kink and antikink waveforms~\cite{katsimiga2023interactions,katsimiga2023solitary,kartashov2022spinor} correspond to Whitham shocks (see Sec.~\ref{LHY-driven-evolution}) formed across the hyperbolic threshold. Remarkably these structures, which are limiting cases of the traveling waves in Fig.~\ref{fig:anatomy}(c), propagate into the vacuum state with a velocity specified by the non-zero homogeneous background. This is in contrast to other scenarios in convex bi-directional dispersive NLS type   hydrodynamics~\cite{el2016dispersive}. 
\begin{table*}[!ht]
\caption{Different types of DSW structures emerging within the eGPE model}\label{tab:notes}
\centering
\begin{tabular}{{|c|c|c|}}\toprule\hline 
Types & Figure & Defining features \\ \hline \hline
Standard DSW & Fig.~\ref{fig:anatomy}(a) & Characterized by soliton (linear) left (right) edge. Exists in the hyperbolic regime \\ \hline DSW Remnant & Fig.~\ref{fig:4}(c) & Similar to standard DSW, but formed across the elliptic-to-hyperbolic threshold \\ \hline Droplet DSW & Fig.~\ref{fig:anatomy} (b) & Contains a nonlinear droplet edge and a linear one, formed within the elliptic regime  \\ \hline Traveling DSW & Fig.~\ref{fig:anatomy} (c) & Composite waveform comprised of a partial DSW and traveling wave segments\\ \hline (Anti)Kink & Fig.~\ref{fig:anatomy} (d) & Robust tanh-type waveforms emerging for  $(\rho_0^{(1)},\rho_0^{(2)})=(4/9,0)$ and $(\rho_0^{(1)},\rho_0^{(2)})=(0,4/9)$ \\ \hline
\end{tabular}
\end{table*}

\begin{figure}
\centering
\includegraphics[width=1.0\linewidth]{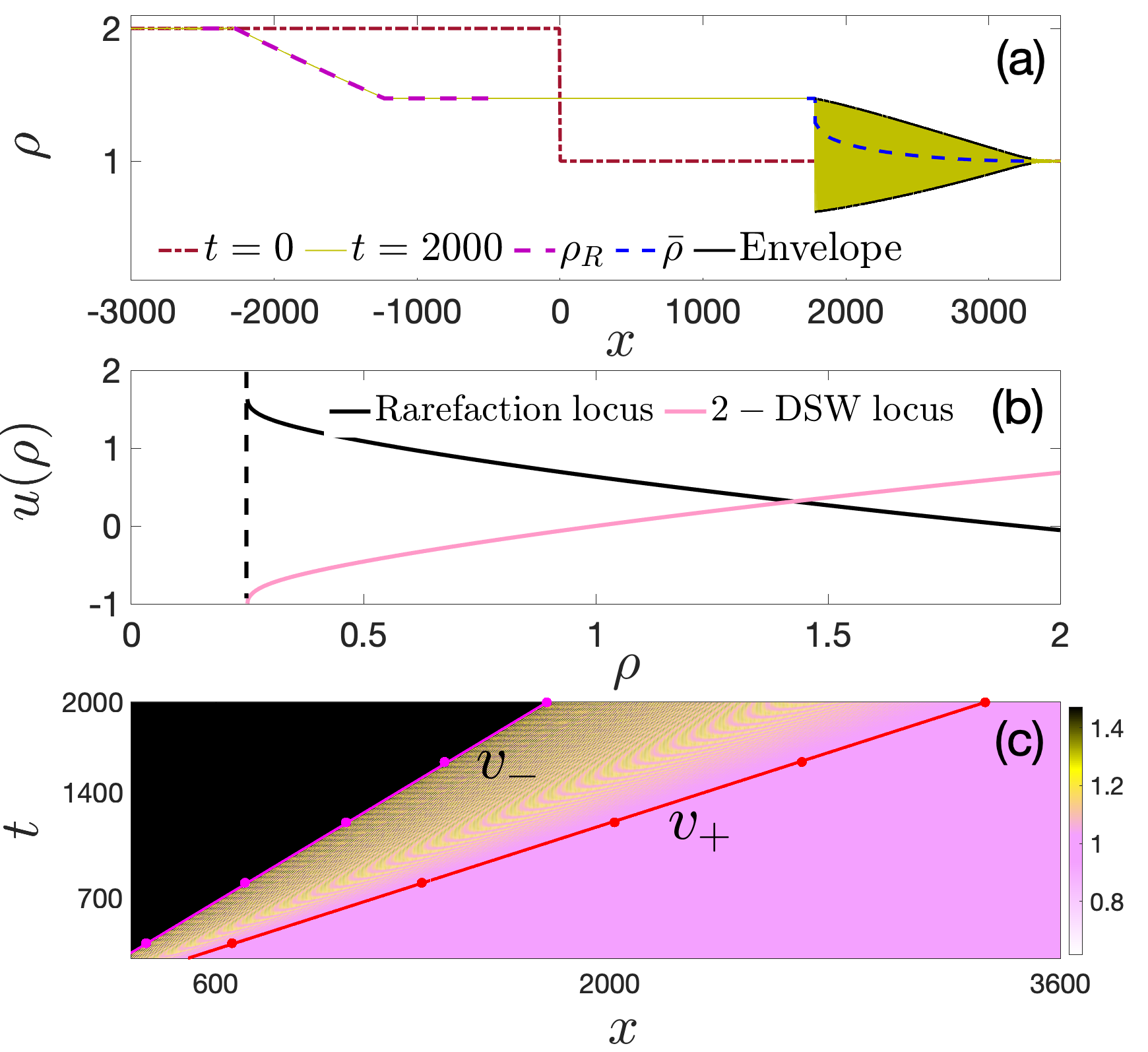}
\caption{(a) Density profile of the initial and long-time evolved (see legend) waveform of a dam break problem for $(\rho_0^{(1)},\rho_0^{(2)})=(2,1)$. 
The analytical rarefaction solution, $\rho_R$, of Eqs.~\eqref{Hydrodynamic-reduction} as well as the  wave mean, $\bar \rho$, and DSW envelopes are also provided (see legend) evincing slow variations across the wavetrain. 
(b) Intermediate hydrodynamic background density and velocity, $(\rho_m,u_m)$, found through the intersection of the 1-rarefaction and the 2-DSW loci. (c) Density evolution using the initial condition of (a) showcasing the generation and propagation of a DSW which occurs in general for $\rho_0^{(1,2)}\gg 0.25$. 
The analytically obtained linear edge, $v_-$, and soliton, $v_+$, speeds are also depicted.}
\label{fig:2}
\end{figure}

\section{Dynamical generation of DSW$\small{\textrm{s}}$}
\label{sec:results}

In the following, we explore the dynamics emanating from the initial condition of Eq.~(\ref{Riemann-step-GP}) (Riemann problem) applied to the eGPE [Eq.~(\ref{GP-eqn})]. 
In particular, the long-time evolution is monitored with an emphasis on the zero velocity, $u_0^{(1)}=0$, scenario also known as dispersive dam break problem~\cite{EL201611}. 
Two distinct regimes are examined based on the existence of the above-discussed hyperbolic-elliptic threshold. 
This investigation corresponds to the variation of $\rho_{0}^{(2)}$ in the initial condition being above (below) the critical value of $\rho_{0}^{(2)}=0.25$, while $\rho_{0}^{(1)}$ is held fixed.  
\begin{figure}
\centering
\includegraphics[width=\linewidth]{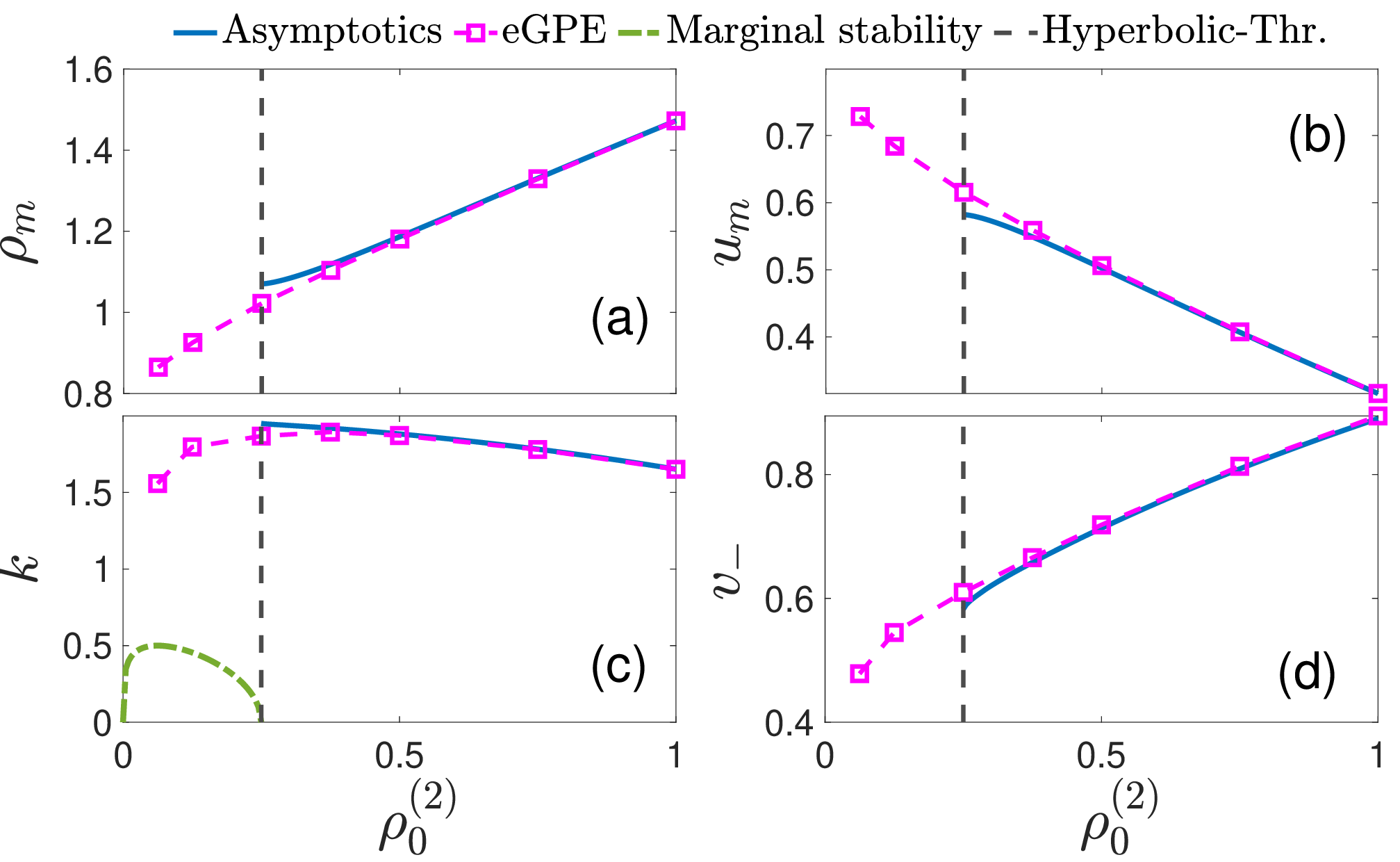}
\caption{Macroscopic characteristics of the  generated DSWs of dam break problems with $\rho_0^{(1)}=2$ and for distinct 
values of $\rho_0^{(2)}$. 
Specifically, the intermediate hydrodynamic (a) density $\rho_m$ and (b) velocity $u_m$ are depicted along with (c) the linear edge wavenumber, $k$, and (d) the soliton edge speed $v_-$. 
The smooth behavior of these quantities across the threshold 
of $\rho_0^{(2)}=0.25$ suggests the persistence of the DSW in a LHY dominated, $\rho_0^{(2)}<0.25$, region. 
Dashed black lines mark the relevant threshold. 
The dashed green line designates the curve of marginal stability below which MI occurs.
In all cases, analytical predictions through an asymptotic procedure (see text) together with the results extracted from the eGPE are illustrated (see legend).} 
\label{fig:3}
\end{figure}

\subsection{{Mean-field driven dynamics with $\rho_0^{(1,2)}\geq 0.25$}}
\label{Mean-field-driven-dynamics}
To ensure that the emergent waves from the  Riemann problem lie above the hyperbolic threshold of the hydrodynamic reduction [c.f. Eq. \eqref{Hydrodynamic-reduction}], we choose as a representative parameter set of the initial wave function that of $\{\rho_0^{(1)}=2,\rho_0^{(2)}=1, u_0^{(1)}=0\}$. 
As such, both asymptotic density states $\rho_0^{(1,2)}\gg 0.25$, see Fig.~\ref{fig:2}(a), implying that the standard mean-field nonlinearity prevails over the LHY one. 
The long-time evolved wave patterns, simulated through the eGPE model, consist of a left propagating rarefaction wave and a right propagating DSW~\cite{hoefer2014shock} interconnected via a plane-wave as shown in Fig.~\ref{fig:2}(a). This plane-wave has an amplitude $\rho_m \approx 1.47$ (see also the discussion below) and velocity $u_m={\rm Im}(\psi_m^{*}(\psi_x)_m)/|\psi_m|^2\approx 0.32$. 
In the context of hydrodynamics, the aforementioned  generation of counter-traveling waves originates from the necessity to equalize the hydrodynamic pressure term [Eq.~(\ref{Quasilinear-system-DH-2})] across the initial step profile [Eq.~(\ref{Riemann-step-GP})]. 
For this reason an expansive (compressive) rarefaction (DSW) wave is produced leaving behind the ``wake" of the new intermediate plane-wave state.  

To analytically estimate the characteristics of this plane-wave for $\rho_0^{(1,2)}>0.25$ an asymptotically ($t\gg 1$) valid Whitham-El closure procedure is used~\cite{el2005resolution,el2005undular,hoefer2014shock,el2007theory}. 
In essence, for the bi-directional dispersive hydrodynamic system of Eqs.~\eqref{Quasilinear-system-DH-1}-\eqref{Quasilinear-system-DH-2}, two families of waves (a rarefaction and a DSW) are anticipated to be emitted when  $u_0^{(1)}=0$. 
Along the left traveling rarefaction wave, alias 1-wave (slower or left), the second Riemann invariant $r_2$ of Eq.~(\ref{Hydrodynamic-reduction})  is constant, while across the DSW (2-DSW) which travels to the right or is faster, the first Riemann invariant $r_1$ is  constant, see also Ref.~\cite{el2005resolution}. 
The process of estimating the properties of the intermediate state is as follows: 
the isoline of $r_2(\rho,u)=r_2(\rho_0^{(1)},u_0^{(1)}) \equiv r_2^{(1)}$ is evaluated and shown in Fig.~\ref{fig:2}(b) alongside with the isoline of  $r_1(\rho,u)=r_1(\rho_0^{(2)},0) \equiv r_1^{(2)}$. 
Their intersection represents the (expected to be formed) intermediate state ($\rho_m,u_m$) in the $u(\rho)-\rho$ phase plane as depicted in Fig.~\ref{fig:2}(b). 
Deviation of the order of $5\%$ is observed between this analytical estimate as compared to our eGPE simulations. 

This agreement further motivates the comparison of the asymptotically valid so-called ``simple" wave solution of the hydrodynamic reduction (Eqs.~\eqref{Hydrodynamic-reduction}) to the expansive left propagating wave profile identified in our simulations, see $\rho_R$ in Fig.~\ref{fig:2}(a). 
Specifically, the internal profile of this simple wave can be obtained through the ansatz $a_1[r_1(x/t),r_2^{(1)}]=x/t$. Moreover, the left and right edge locations of the rarefaction wave for a fixed time-instant can be determined via $x=a_1(\rho_0^{(1)}, u_0=0) t$ and $x=a_1(\rho_m,u_m) t$ respectively.  
Fig.~\ref{fig:2}(a) illustrates the excellent agreement between the analytically ascertained self-similar density  profile and the one extracted from our simulations exemplarily at $t=2000$.

Next, we track the emergent DSW propagating towards the right while spreading in the course of the evolution. 
The DSW is a multiscale periodic waveform having an internal microstructure, together with distinct gray solitonic and linear edges. 
Fig.~\ref{fig:2}(a) presents the slow variation of the wave mean $\overline{\rho}$ across the DSW from $\rho_m$ (solitonic edge) to $\rho_0^{(2)}$ (linear edge). 

Here, in order to track the edge speeds of the emitted DSW, we use a reduced-order method based on the system of Whitham modulation equations~\cite{el2005resolution,hoefer2014shock}. 
Three of them for Eq.~\eqref{GP-eqn} can be obtained by averaging its underlying system of ``local" conservation laws given by  Eqs.~\eqref{mass}. This averaging is performed over the rapid oscillations of the family of the eGPE periodic waveforms characterized by four physical parameters. Namely amplitude $a$, wavenumber $\kappa$, mean velocity $\bar u$, and mean density $\bar \rho$. 
Specifically, three out of four Whitham modulation equations in DH form, utilized herein, read 
\begin{subequations}
\begin{align}
\label{rho-equation}
&\overline{\rho}_t+\overline{(\rho u)}_x=0,\\
\label{momentum-equation}
&\overline{\rho u}_t+\overline{\bigg(\rho u^2+\mathsf{P}(\rho)-\frac{\rho}{4}({\rm ln}(\rho))_{xx}\bigg)_x}=0,\\
\label{Hamiltonian-equation}
&\overline{\bigg(\frac{\rho_x^2}{8\rho}+\frac{\rho u^2}{2}+\frac{\rho^2}{2}-\frac{2\rho^{3/2}}{3}\bigg)_t}+\\\nonumber&\frac{1}{2}\overline{\bigg(\rho u^3+2\rho(\rho-\sqrt{\rho})u+\frac{\rho_x^2}{4\rho}u-\frac{\rho_{xx}u}{2}+\frac{\rho_x(\rho u)_x}{2\rho}\bigg)}_x=0. 
\end{align} \label{Whitham-set1}
\end{subequations} 
In all cases, the overbar indicates averaging of the associated terms over an oscillation period. 
A fourth modulation equation, the so-called ``conservation of waves", $\kappa_t+\omega_x=0$ is needed to describe the slow variation of the additional  parameter (wavenumber $\kappa$), thus closing the system. 
All Whitham are valid for $x,t\gg1$. 
\begin{figure}
\includegraphics[width=0.48\textwidth]{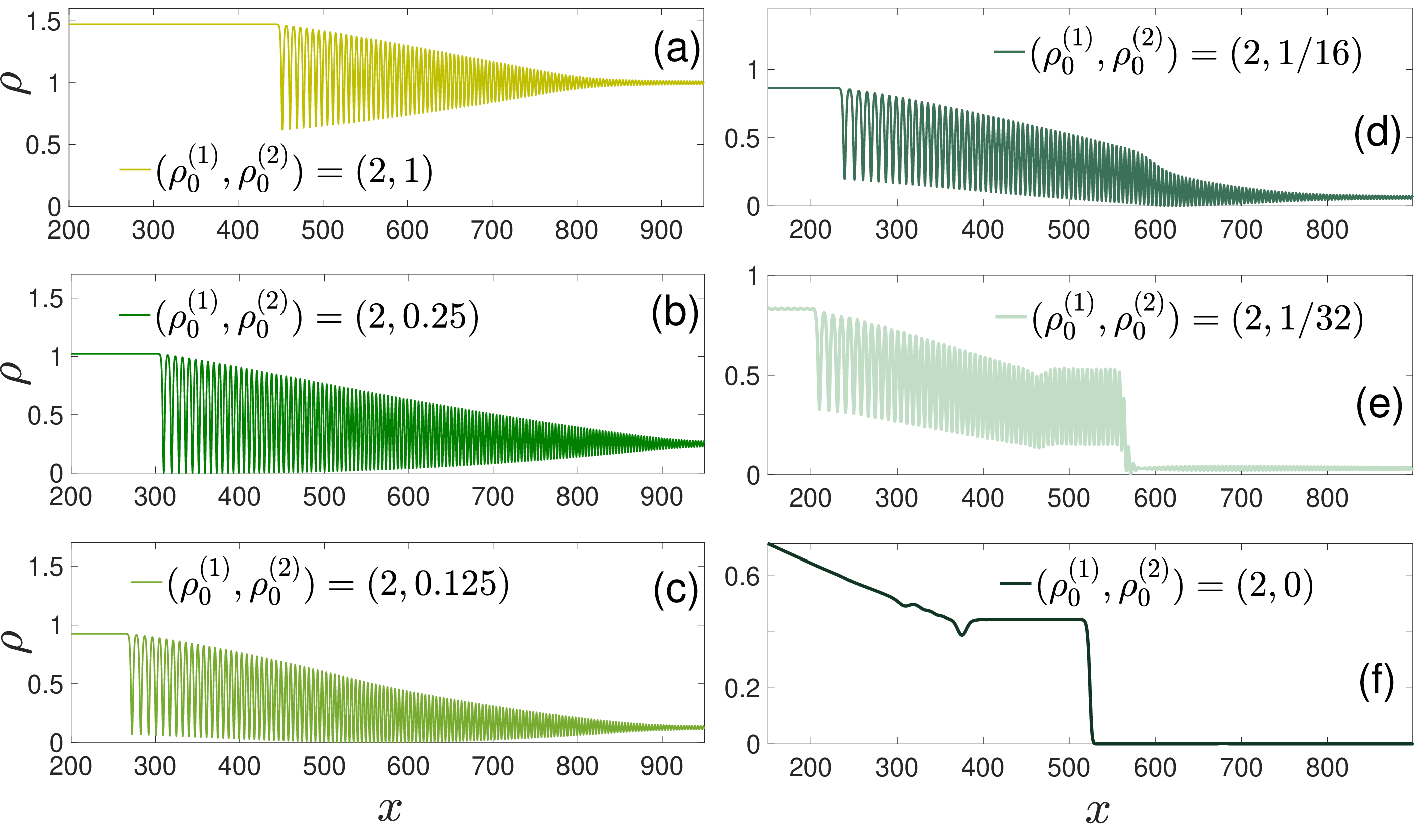}
\caption{Density snapshots at $t=500$ depicting the emitted right propagating waves in the case of a dam break problem with fixed $\rho_0^{(1)}=2$ and varying $\rho_0^{(2)}$ (see legends). 
DSWs persist across the hyperbolic threshold [panels (a)-(d)] transitioning thereafter to  traveling DSWs [panel (e)] and finally in the vacuum case to an antikink entity [panel (f)].} 
\label{fig:4}
\end{figure}

The spatiotemporal evolution of the density of the dam-break problem with $(\rho_0^{(1)}, \rho_0^{(2)})=(2,1)$ is depicted in Fig.~\ref{fig:2}(c).  
As it can be seen, a DSW emerges characterized by an upstream (solitonic) and a downstream (linear) front that are associated with two distinct velocities, namely $v_-$ and $v_+$. 
These velocities can also be analytically predicted deploying the aforementioned Whitham modulation equations, see Appendix~\ref{appendix:Whitham_eq}. 
Specifically, the expression for the speed of the linear edge of the 2-DSW reads 
\begin{equation}
 v_{+}(\rho_0^{(2)})= \bigg((2c_0^{(2)}(\alpha_0^{(2)})^2-c_0^{(2)})/\alpha_0^{(2)}\bigg),
\end{equation}
where $\alpha_0^{(2)}=\sqrt{k^2/(4[c_0^{(2)}]^2)+1}$ signifies the scaled phase speed of the linear edge.  
Moreover, the wavenumber at the linear edge reads
\begin{equation}
\label{wavenumber-linear-edge_main}
k(\rho_0^{(2)})=2(c(\rho_0^{(2)}))\sqrt{(\alpha(\rho_0^{(2)}))^2-1}.
\end{equation} 
\begin{figure}
\includegraphics[width=\linewidth]{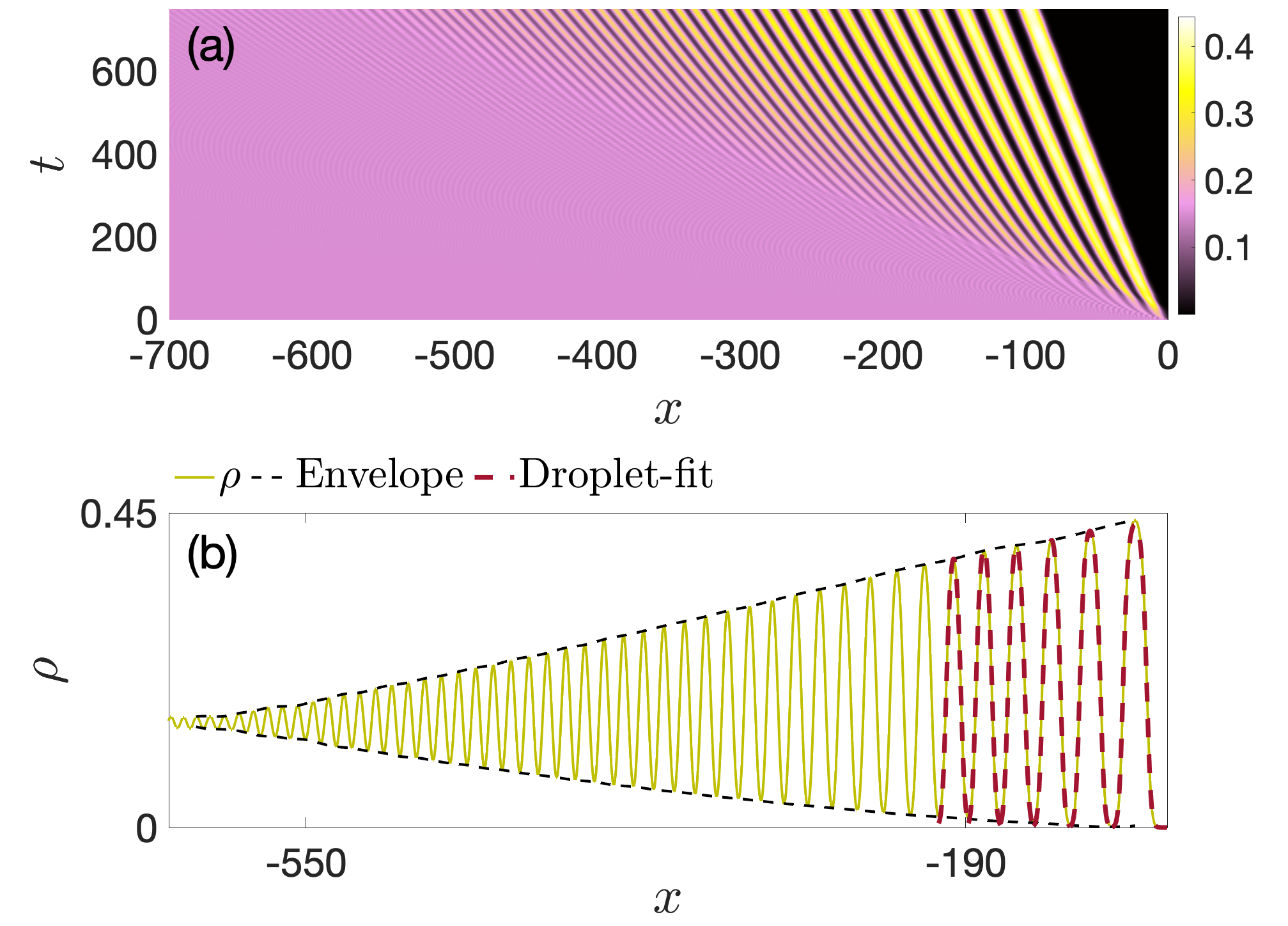}
\caption{(a) Spatiotemporal density evolution of a nucleated droplet DSW for a vacuum dam break problem characterized by ($\rho_0^{(2)},\rho_0^{(1)})=(0,0.15$). Its formation originates from the MI, and such a wave develops and propagates robustly  for long evolution times. 
(b) Density profile of the ensuing droplet train at $t=750$. 
Dashed black lines indicate the envelope of the wave pattern, whereas dashed red lines correspond to six fitted exact droplet solutions of the eGPE (see text).}
   \label{fig:6}
\end{figure}
On the other hand, the conjugate phase speed (see the relevant definition in the  Appendix~\ref{appendix:Whitham_eq}) at the soliton edge of the 2-DSW yields the soliton phase speed
\begin{equation}
v_{-}=u_m+c_m\sqrt{1-\frac{\Tilde{k}_m^2}{4c_m^2}},\label{soliton_edge_speed}
\end{equation}
where $c_m=c(\rho_m)$ is the local speed of sound at the intermediate state and $\Tilde{k}_m$ is the conjugate wavenumber at the soliton edge which is essentially analogous to the wave amplitude. 
These analytically predicted velocities for both the soliton, $x/t=v_{-}$, and the linear, $x/t=v_{+}$, edges are illustrated in Fig.~\ref{fig:2}(c). 
Evidently, an  excellent agreement between the analytics and the eGPE predictions is observed.

As a next step, the family of dam break problems obtained by fixing $\rho_0^{(1)}=2$ and varying $0.25\leq \rho_0^{(2)}<\rho_0^{(1)}$ is investigated. 
The variation of the four variables characterizing the macroscopic properties of the generated 2-DSW as a function of $\rho_0^{(2)}$, is displayed in Figs.~\ref{fig:3}(a)-(d). 
Concretely, these variables refer to 
the intermediate density ($\rho_m$) and velocity ($u_m$), the linear edge wavenumber ($k(\rho_0^{(2)})$) and the soliton edge speed $v_{-}$. 
The results extracted from the eGPE simulations are shown as magenta squares, against the asymptotic curves (blue solid lines). The latter are ascertained using Eqs.~\eqref{wavenumber-linear-edge_main}, \eqref{soliton_edge_speed} and by estimating the intermediate state through the intersection of the isolines of the relevant Riemann invariants (see e.g. Fig.~\ref{fig:2}(b)). {The intermediate density $\rho_m$ and velocity $u_m$ are identified at later evolution times ($t\gtrapprox500$) from the relevant snapshots. The linear (soliton) edge velocities $v_+=\partial_k\omega_0$ ($v_-$) are extracted by fitting a straight line to either of the edges in the spatiotemporal density evolution.
{Moreover to obtain the linear edge wavenumber we invert the velocity $\partial_k\omega_0$. Despite the multivaluedness therein, we observe $k(\rho_0^{(2)})>k_i$, where $\partial_{kk}\omega_0(k_i)=0$.} 
Excellent agreement between numerical and analytical (asymptotic curves) findings is witnessed, with the observed deviations for $\rho_0^{(2)}\lessapprox 0.5$ 
being typically less than $5\%$.}

It turns out that this deviation can be interconnected with the role of the LHY contribution already within the hyperbolic regime. 
{Recall} that the hydrodynamic pressure curve $P(\rho)$, possesses a root at $\rho_0^{(2)}=4/9$.   
As a consequence, there exists an interval of densities in the hyperbolic regime, i.e. $0.25\leq \rho_0^{(2)}<4/9$, for which  $\mathsf{P}(\rho_0^{(2)})<0$. 
{A negative pressure region that would be absent without the LHY contribution.} 
Intuitively, the flow resulting from the dam break problem is driven by the pressure difference $\Delta \mathsf{P}=\mathsf{P}(\rho_0^{(1)})-\mathsf{P}(\rho_0^{(2)})$.  {If $\mathsf{P}(\rho_0^{(2)})<0$,  $\Delta \mathsf{P}>\mathsf{P}(\rho_0^{(1)})$ is enhanced which leads to an increased linear momentum growth rate     
$\frac{d}{dt}\int_{-\infty}^{\infty}\rho u dx =\Delta \mathsf{P}=\mathsf{P}(\rho_0^{(1)})-\mathsf{P}(\rho_0^{(2)})$.}

Representative density profiles of the emitted 2-DSW for $\rho_0^{(1)}=2$ and varying $0.25\leq \rho_0^{(2)}<\rho_0^{(1)}$ are presented in Figs.~\ref{fig:4}(a), (b). 
Here, the focus is the deformation of DSW patterns across the threshold. 
Within this $\rho_0^{(2)}$ interval there is a gradual slowdown of the soliton speed, relative to the background given by $c_s=v_--u_m$. 
The latter is a generic feature of Schr\"odinger-type systems~\cite{EL201611}. For $\rho_0^{(2)}=0.25$ a black soliton edge  occurs [Fig.~\ref{fig:4}(b)] resulting in a cavitation point, i.e. a point of zero density therein. 
Here, we should also emphasize the longevity of the 2-DSW waveforms for  $0.25\leq\rho_0^{(2)}<2$, as they develop coherently across  very long simulation times, $t\sim 2000$.

\subsection{{LHY driven evolution with $\rho_0^{(1)}\leq 0.25$ or $\rho_0^{(2)}<0.25$ }}
\label{LHY-driven-evolution}

Subsequently, we extend 
the family of dam break problems into the elliptic regime where $\rho_0^{(1)}=2$ and $\rho_0^{(2)}$ is reduced below the hyperbolic threshold, i.e. $\rho_0^{(2)}<0.25$, see also  Sec.~\ref{hydrodynamics}. 
At threshold, the speed of sound $c(\rho_0^{(2)})=0$ and thus both Riemann invariants (and the relevant hyperbolic speeds $a_{1,2}$) of Eq.~\eqref{Hydrodynamic-reduction} coincide. 
Below the  threshold, the associated speed of sound is imaginary  and as a consequence, the Whitham modulation equations become elliptic (c.f. Eqs.~\eqref{density-lin-edge},\eqref{velocity-lin-edge}).
This ellipticity impacts the existence, stability, and lifetimes of the emitted DSW. 
{Generally, in attractive environments, DSWs can be observed only under very specific circumstances \cite{EL201611,el2016dam,biondini2018riemann}}. However, in our case as we explicate below, the eGPE admits robust DSW generation. 

Indeed, as it can be seen in Figs.~\ref{fig:4}(c), (d), even for $\rho_0^{(2)}<0.25$  development of coherent {\it ``DSW-remnants"} takes  place up to $t \sim 500$.  
The parametric variation of the macroscopic properties $\rho_m$, $u_m$, $k$, and $v_{-}$ of these DSW remnants is depicted in Figs.~\ref{fig:3}(a)-(d). 
These observables exhibit a smooth transition across the hyperbolic threshold into the elliptic regime. They demonstrate that the aforementioned remnants represent an extension of the DSW into this regime since they connect a stable [$\rho_m>0.25$, c.f. Fig.~\ref{fig:3}(a)] and an unstable ($\rho_0^{(2)}<0.25$) hydrodynamic state. Specifically, $\rho_m$ ($u_m$) decreases (increases) across the threshold, whilst $k$ (at the linear edge) reduces accompanied by a corresponding descent of the linear edge speed. Importantly, $k$ lies outside the interval of unstable wavenumbers $0<k<k_c$, with the threshold of marginal stability being defined as   $k_c(\rho_0)=\sqrt{2\sqrt{\rho_0}-4\rho_0}$. 
In the elliptic regime, besides a small band of unstable wavenumbers $0<k< \sqrt{2\sqrt{\rho_0}-4\rho_0}$, shorter waves are regularized by the underlying dispersion in the DH Eqs.~\eqref{Quasilinear-system-DH-1}-\eqref{Quasilinear-system-DH-2}. 
{The aforementioned stabilization mechanism hints towards the longevity of these ``remnants". A similar, but less pronounced, stabilization scenario manifests in the case of the classical, attractive cubic NLS equation \cite{gurevich1993modulational,biondini2016universal,biondini2018riemann,gurevich1993modulational}.}

For longer evolution times (not shown), $t\gtrapprox 600$, the {envelope of the} DSW remnant begins to break down {in contrast to the standard DSW that does not experience such a breakdown}.
The associated instability develops in the vicinity of the DSW linear edge and propagates into its interior. 
This indicates that the eventual breakdown of the remnant is solely a consequence of MI associated with $\rho_0^{(2)}$. 
An estimate for the breakdown time can be extracted by assuming exponential growth of the instability with its associated rate\red{\footnote{{The characteristic time of destabilization is given by 
$t_c\sim \frac{2}{k_{{\rm max}}^2}{\rm ln}(\frac{\sqrt{\rho_0}}{\epsilon}), \;\epsilon \ll 1$ where $\epsilon$ is a representative perturbation amplitude.}}} given by the amplitude of the mode of maximal growth, at $k_{{\rm max}}=\sqrt{\sqrt{\rho_0}-2\rho_0}$, in the spectral sideband at $0<k<k_c$. A similar degree of destabilization has been reported for attractive dispersive dam-break flows in cubic NLS~\cite{gurevich1993modulational,el2016dam,wan2010diffraction}.

\begin{figure}
\includegraphics[width=0.48\textwidth]{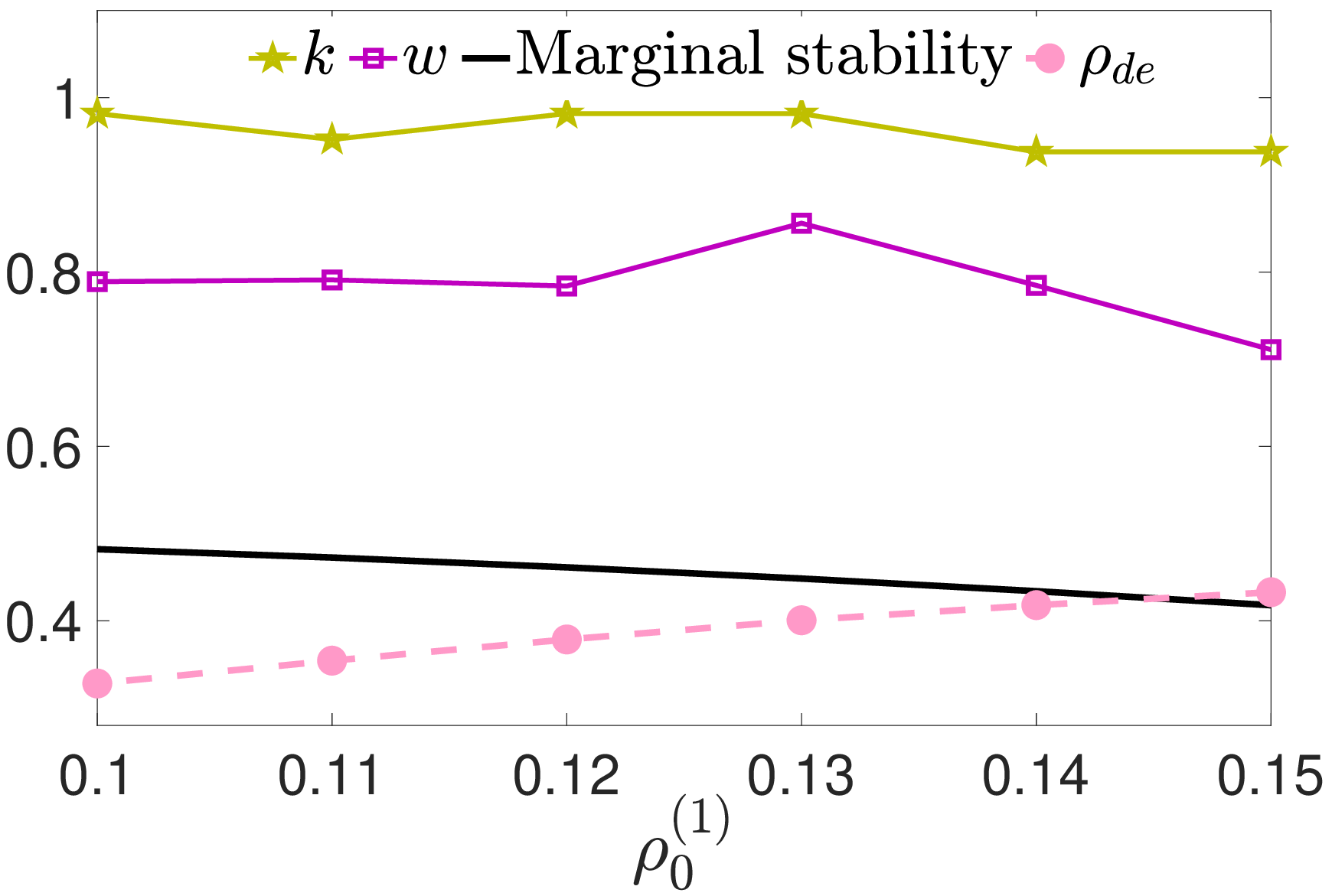}
\caption{Properties of the emitted droplet DSWs throughout their $\rho_0^{(1)}$ domain of existence. 
The wavenumber, $k$, at the linear edge of the droplet DSW displays weak variations and lies above the curve of marginal stability (see legend). The relevant width, $w=v_{+}-v_{-}$, of the droplet DSW fluctuates in the interval $w\in (0.7, 0.9)$. Additionally, the droplet edge height, $\rho_{de}$, as a function of $\rho_0^{(2)}$, shows a monotonic increase before its saturation at $\lessapprox 4/9$.}
\label{fig:7}
\end{figure}

Closely inspecting Figs.~\ref{fig:4}(b)-(d) it becomes evident that the location of the cavitation point of the DSW shifts into its interior as $\rho_0^{(2)}$ is reduced. Furthermore, the envelope of the DSW corresponding to $\rho_0^{(2)}=1/16$ [Fig.~\ref{fig:4}(d)] displays a prominent curvature near its linear edge. 
These observations motivate the exploration of the nucleated structures for decreasing $\rho_0^{(2)} \to 0$, see Figs.~\ref{fig:4}(e), (f). 
For sufficiently small $\rho_0^{(2)}$ as shown in Fig.~\ref{fig:4}(e) and Fig.~\ref{fig:anatomy}(c), the DSW ``remnant" transitions to a composite waveform characterized by two segments. 
The first segment refers to a partial DSW, which connects the intermediate hydrodynamic background $(\rho_m, u_m)$ to an eGPE periodic wave possessing edge speeds $v_{-} $ and $v_p$ respectively. Due to the distinct modulation speeds at its edges, the partial DSW spreads as it propagates.
The second segment however, is a waveform that travels with speed $v_s$, linking the aforementioned periodic wave and the hydrodynamic background $\rho_0^{(2)}$ across very few density oscillations [akin to c.f.~\cite{sprenger2017shock}]. Such composite waveforms comprising of partial DSW and traveling wave segments have been referred to as {\it``traveling DSW"} in their observations in shallow water waves in which higher-order nonconvex dispersive effects are prevalent \cite{hoefer2019modulation,sprenger2017shock,sprenger2019generalized,sprenger2023traveling}. 
We have checked that such {traveling DSW} occur within the interval $0<\rho_0^{(2)}\leq {1}/{32}$. 
{Similarly to the DSW remnants, these waveforms are long-lived up to $t\gtrapprox 600$ and their eventual breakdown appears to be due to the MI of $\rho_0^{(2)}$. 
Interestingly, the traveling {segment} satisfies the Rankine-Hugoniot jump conditions \cite{leveque2002finite} $\Vec{F}(\Vec{U}_L)-\Vec{F}(\Vec{U}_R)=v_s(\Vec{U}_L-\Vec{U}_R)$ of the system of Eqs.~\eqref{Whitham-set1} written in the compact form 
$\vec{{U}}_t+(\vec{{F}}(\vec{U}))_x=0$, where the subscript $L$ ($R$) stands for the left periodic (right hydrodynamic background $\rho_0^{(2)}$) states. At the level of the modulation equations, $\Vec{U}_{L,R}$ represent the averaged quantities, which for a periodic (or homogeneous background) wave solution correspond to constant parameters. Thus, these novel waveforms appear to be {\it ``Whitham-shocks"}, i.e., shock solutions of the Whitham modulation equations of the eGPE. 
While we are not aware of previous reports of such structures in the present setting,
Whitham-shocks have been extensively studied in the context of the Kawahara equation modeling shallow surface water waves incorporating higher-order nonconvex dispersion \cite{sprenger2020discontinuous,sprenger2017shock,sprenger2023traveling}. }

\begin{figure}
\includegraphics[width=0.48\textwidth]{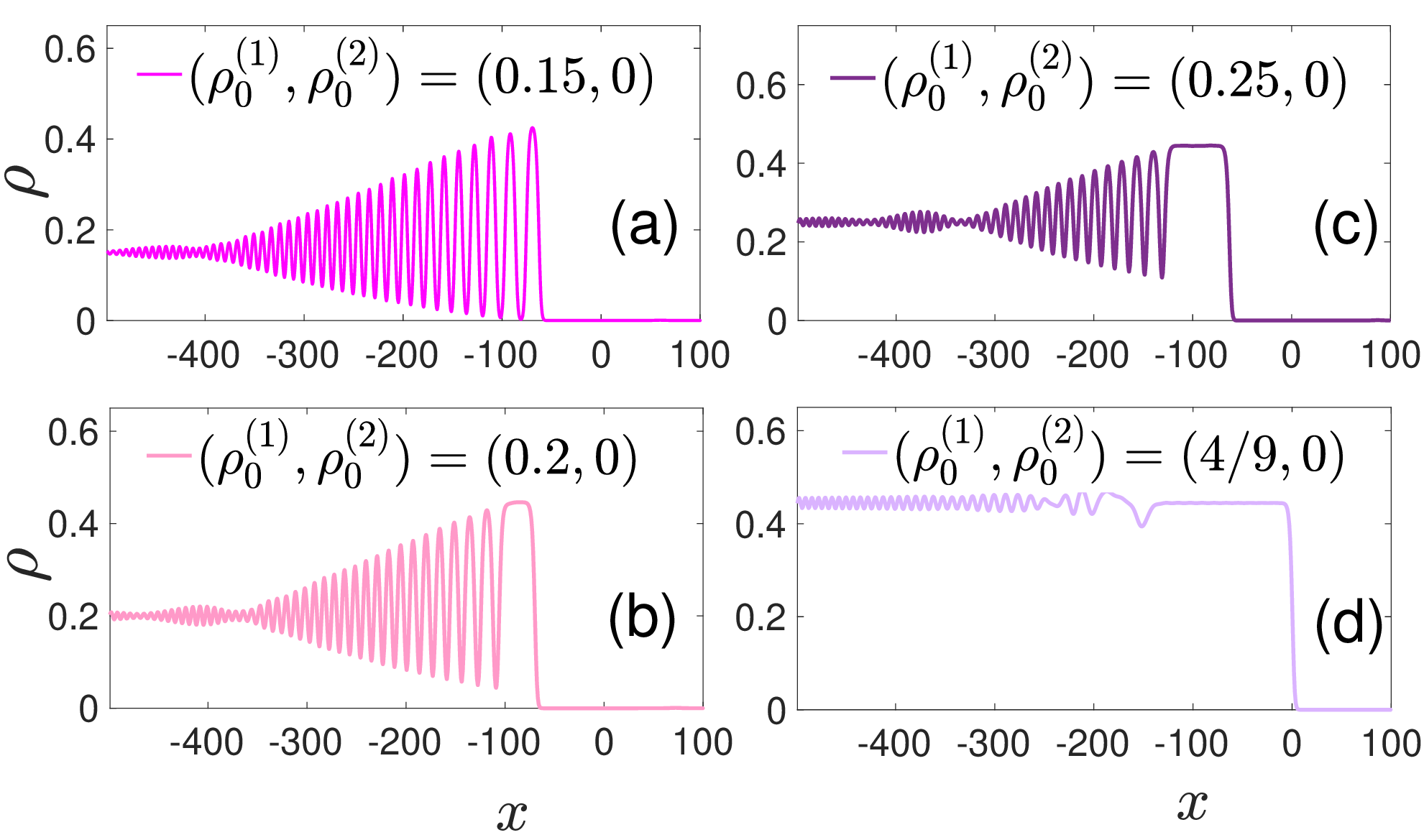}
\caption{Density profiles of the wave patterns at $t=500$ emerging from the class of vacuum dam break problems with $0<\rho_0^{(1)}\leq 4/9$. (a) Droplet DSW arising for $\rho_0^{(1)}\leq 0.15$ (see legend).  
(b) Turning to $\rho_0^{(1)}>0.15$, the right droplet edge transitions to an antikink which is  connected to the homogeneous background $\rho_0^{(1)}$ through a 1-DSW remnant. 
(c) Same as (b) but with a repulsive 1-DSW connecting the antikink to the respective  background.  
(d) Considering $\rho_0^{(1)}=4/9$, a non-moving  antikink emerges corresponding to zero pressure difference of the underlying dam break problem.}
\label{fig:5}
\end{figure}
Turning to the case of $\rho_0^{(2)}=0$, i.e., the vacuum dam break Riemann problem, we observe that 
the intermediate hydrodynamic density is $\rho_m =4/9$, see Fig.~\ref{fig:4}(f). 
Here, the right traveling waveform transitions to an antikink configuration which is an exact, stable solution to the eGPE~\cite{katsimiga2023interactions}. This stability translates to the robust evolution of the entire structure monitored for times up to $t=2000$. 
Remarkably, it turns out that the antikink is a dispersive hydrodynamic structure that spontaneously emerges from a family of vacuum ($\rho_0^{(2)}=0$) dam break problems having $\rho_0^{(1)}> 0.15$ and propagating with $u_m$. \footnote{{For $\rho_0^{(1)}>4/9$, an estimate for $u_m$ is analytically given by $\int_{4/9}^{\rho_0^{(1)}}\frac{c}{\rho}d\rho$ (1-wave condition) with the overall pattern comprising of a left (right) propagating rarefaction wave (antikink).}} 
{It turns out that} the antikink (and kink) itself is a Whitham shock {traveling with $u_m$, and} satisfying the Rankine-Hugoniot jump conditions of Eqs.~\eqref{rho-equation}-\eqref{Hamiltonian-equation} {since it connects} two homogeneous plane wave backgrounds characterized by zero hydrodynamic pressure, i.e. $\mathsf{P}(4/9)=\mathsf{P}(0)$. 

As a next step, we maintain  $\rho_0^{(2)}=0$ but vary  $\rho_0^{(1)}\in (0,4/9]$.
In this case, since $\Delta \mathsf{P}=\mathsf{P}(\rho_0^{(1)})<0$, the growth rate of the integrated momentum is negative, {i.e. $\frac{d}{dt}\int_{-\infty}^{\infty} \rho u dx<0$ }, suggesting that the emitted pattern propagates to the left. Indeed, monitoring the density evolution shown in Fig.~\ref{fig:6}(a) for $\rho_0^{(1)}=0.15$, we observe the emission of a one-phase modulated wavetrain, which remains robust up to $t=750$. Interestingly, the density  oscillations in this wavetrain correspond to droplets, as can be seen from Fig.~\ref{fig:6}(b),  which provides a fitting of the analytical 1D droplet (and stable) solution, [see Eq.~\eqref{Droplet-solution}], for the first six waves in the wavetrain. 
We dub this coherent and robust wavetrain as {\it droplet ``DSW"}. The longevity of this entity is hinted from the known droplet stability~\cite{katsimiga2023solitary}, with its  eventual breakdown occurring seemingly due to the MI of $\rho_0^{(1)}$ for $t\gtrapprox 800$. Here, the filamentation of the droplet DSW is initiated at the linear edge, as was the case also for the DSW remnants [Fig.~\ref{fig:4}(d)].

However, in contrast to DSW remnants, the mechanism for the initiation of droplet DSWs is the dominant presence of the attractive LHY term. 
The anti-hydrodynamic mechanism, entailed by negative pressure and imaginary speed of sound \cite{gurevich1993modulational} is in sharp contrast to the formation of remnants and standard DSWs developing across and above the hyperbolic threshold. Indeed,  their generation is related to the prevention of wave breaking being the standard way of formation in fluid-like media.   
We remark that the aforementioned droplet DSW can also be  produced for $\rho_0^{(1)}\leq 0.15$. Similar robust, yet bright solitonic wavetrains are known to emerge in the vacuum dam break problem in the context of the cubic attractive NLS (and nonintegrable variants thereof), studying the universal stage of MI {\cite{biondini2016universal,gurevich1993modulational,biondini2018universal}} and rogue wave generation \cite{el2016dam}.

The macroscopic properties of this one-parameter droplet DSW family ($\rho_0^{(1)}\leq 0.15$) are presented in Fig.~\ref{fig:7}. 
Notice that $k$ at the linear edge fluctuates around unity as a function of $\rho_0^{(1)}$. However, $k(\rho_0^{(1)}) \approx 1$ lies outside the band of unstable wavenumbers bounded by the curve of marginal stability, $k_c(\rho_0)$, illustrated in Fig.~\ref{fig:7} with solid black line. 
The width, $w$, of the droplet DSW is found to display insignificant variation with $\rho_0^{(1)}$.  
Furthermore, the behavior of the peak density [$\rho_{\rm de}(\rho_0^{(1)})$] of the droplet edge features a linear growth for $\rho_0^{(1)}\ll 0.15$ which begins to saturate to a value less than $4/9$. 

For $0.15<\rho_0^{(1)}\leq 4/9$ and $\rho_0^{(2)}=0$, droplet DSWs are no longer present. 
Instead, the right droplet edge is seen to {deform} into an antikink entity, see for instance Fig.~\ref{fig:5}(b) where $\rho_0^{(1)}=0.2$. 
This difference between the left background density of the antikink and $\rho_0^{(1)}$ produces a 1-DSW remnant which acts as the regularization mechanism. 
For a larger $\rho_0^{(1)}$ such as $0.25\leq \rho_0^{(1)}<4/9$ depicted in Fig.~\ref{fig:5}(c), repulsive 1-DSW wavetrains nucleate being stable for adequately long evolution times  $t=2000$. 
Moreover here, it is possible to extract the intermediate background velocity $u_m<0$ (and thus the velocity of the antikink) by utilizing the 1-DSW jump condition. 
This prediction is found to be in agreement within $5\%$ with the eGPE simulations. 
Further increasing the value of the left background to $\rho_0^{(1)}=4/9$, it holds that $\Delta \mathsf{P}=\mathsf{P}(4/9)=0$ implying a zero momentum growth rate. This explains the dynamical formation of a non-moving antikink, as can be attested from the eGPE simulations presented in Fig.~\ref{fig:5}(d). 
Notice also the existence of small amplitude undulations in the non-zero background caused by the discrepancy between the width of the initial condition as compared to that of the antikink. 

\begin{figure}
\centering
\includegraphics[width=0.51\textwidth]{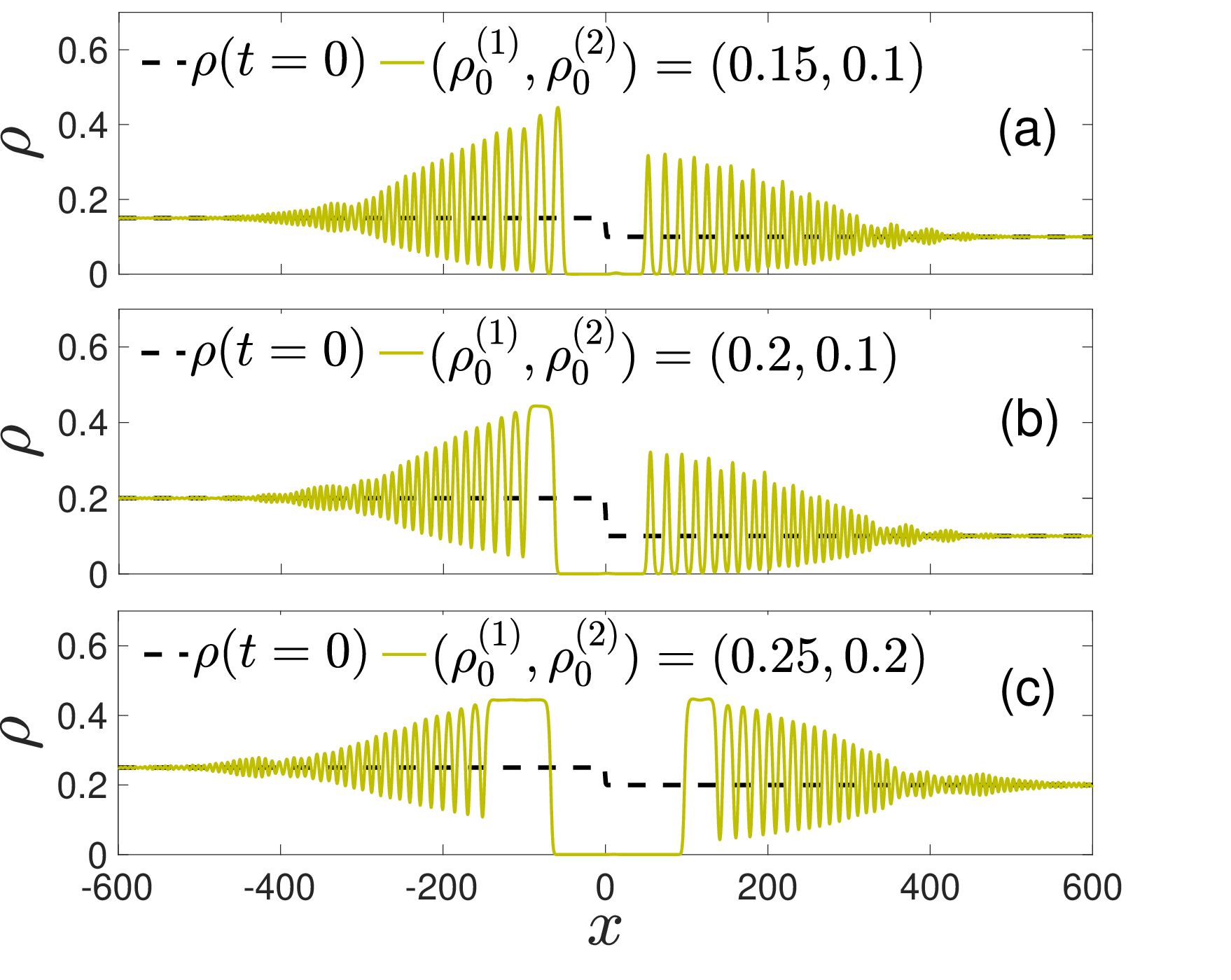}
\caption{Density snapshots of counterpropagating (a) droplet DSW, (b) antikink-droplet DSW both at $t=500$, and (c) antikink-kink pairs at $t=750$. 
The latter are connected to the external ``flow" via 1-DSW and 2-DSW remnant respectively. 
In all cases, the emergent configurations occur below the hyperbolic threshold and appear to be ``superpositions" of vacuum dam break Riemann problems with ($\rho_0^{(1)},0$) and ($0,\rho_0^{(2)}$). }
\label{fig:8}
\end{figure}

The final category of dam break problems corresponds to {$\rho_0^{(1)}< 0.25$} and $\rho_0^{(2)}\neq 0$. Here, the speed of sound associated with each homogeneous background is complex, thus leading to a regime where effects quite contrary to fluid-like behavior dominate. 
{Three} characteristic examples are illustrated in Figs.~\ref{fig:8}(a)-(c) 
 {referring to $(\rho_0^{(1)},\rho_0^{(2)})= (0.15,0.1)$, $(\rho_0^{(1)},\rho_0^{(2)})= (0.2,0.1)$, and $(\rho_0^{(1)},\rho_0^{(2)})= (0.25,0.2)$, respectively}. 
Evidently, in the former case two counterpropagating droplet DSWs occur [Figs.~\ref{fig:8}(a)], while in the second scenario, the emergent counterpropagating flows consist of a left-traveling antikink and 1-DSW remnant together with a right-traveling droplet DSW [Figs.~\ref{fig:8}(b)]. 
On the other hand, by setting $\rho_0^{(1,2)}>0.15$ (or generally above the threshold of the existence of droplet DSWs), one can realize the formation of a counterpropagating antikink-kink pair [Figs.~\ref{fig:8}(c)]. 

We remark that the piecewise constant initial condition (Riemann problem) can be viewed as a superposition of two different vacuum dam-break problems. Astonishingly, it turns out that the resulting wave patterns here emerge from the individual vacuum dam break problems with 
($\rho_0^{(1)},0$) (having $0<\rho_0^{(1)}<0.25$) and ($0,\rho_0^{(2)}$). 
A similar response has been reported for Riemann problems in the absence of the LHY, but for an attractive NLS model~\cite{biondini2018riemann}. However, in our case, such a superposition aspect could also lead to kink-antikink pairs, in sharp contrast to the NLS scenario. This is primarily related to the manifestation of the underlying regularization mechanism transitioning from the droplet DSW [Fig.~\ref{fig:5}(a)] to an antikink together with the upstream 1-DSW [Fig.~\ref{fig:5}(b)] when increasing $\rho_0^{(1)}>0.15$. However, in the attractive NLS, it always corresponds to bright solitonic wavetrains \cite{biondini2016universal,biondini2018riemann,gurevich1993modulational}. This is somewhat natural to expect as the absence of
competing nonlinearities does not enable the existence
of a kink or antikink solution in the latter setting.
%
\section{Summary and future challenges}\label{conclusions}

We explored the on-demand dynamical generation of a plethora of dispersive shock waves arising in attractive homonuclear mixtures, as captured by the 1D eGPE model.
Particularly, by designing step-like initial conditions characterized by two densities and a relative velocity (being set to zero) of the underlying Riemann problem utilized herein, we were able to identify a
threshold that designates distinct dynamical response regimes.
This threshold, existing due to the interplay between standard mean-field repulsion (hyperbolic) and LHY attraction (elliptic), separates regions of real-valued vs imaginary speed of sound. 
This is analytically obtained via the
hydrodynamic reduction of the eGPE. 

In the former situation, where mean-field interactions prevail, rarefaction and DSW nucleation is evidenced with the two entities being interconnected through an intermediate plane wave.
Full characterization of the emitted patterns is provided utilizing the Whitam-El method  describing, for instance, the rarefaction wave profiles as well as the velocity and density of the intermediate background or the edge speed of the DSW. 
Overall, our analytical predictions are in excellent agreement with the relevant eGPE simulations.   

Remarkably, in the elliptic regime, where the  speed of sound becomes imaginary and thus MI is present, robust DSW nucleation also occurs.
Specifically here, we identify the formation of:
(i) DSW-remnants (namely DSW but unexpectedly within the elliptic regime). 
These transition to (ii) composite traveling DSW, and (iii) finally to an antikink structure.
The latter, emerges as a limiting case of the ensuing dam-break problem and it is found for the first time herein, to be also a member of the family of Whitham-shocks. 
(iv) Droplet-DSWs are also identified in this regime, referring to long-lived shocks that bear in their trailing edge droplets. 
We were able to obtain the cutoff of formation of these waveforms, in terms of nonzero hydrodynamic background and also their deformation towards
antikink-kink pairs connected with 1-DSW and 2-DSW remnants respectively. 
Notably, all of the aforementioned dispersive entities owe their nucleation to the presence of the LHY contribution.

There is a multitude of future research directions emanating from our present findings. 
Chiefly amongst them, is the systematic investigation of the existence and stability properties of periodic and solitary traveling waves~\cite{katsimiga2023solitary}. 
{This can be partially tackled by means of phase-plane analysis and by employing Newton-like iterative schemes~\cite{yang2009newton,kelley1995iterative}. The characterization of stability of periodic waves can be done via (a) Floquet theory and (b) investigation of MI to the full Whitham modulation equations, to which periodic waves are homogeneous backgrounds (c.f. \cite{maiden2016modulations}).
Moreover, an analytical characterization of the (anti)-kink to (1)2-DSW patterns identified herein by self-similar solutions to the Whitham modulation equations in the elliptic regime would be intriguing and should be of relevance to other generalized NLS type equations, 
such as, e.g., the cubic-quintic NLS~\cite{lee2004quantum,crosta2012crossover,crosta2011bistability,araujo}. }

Two other related, and compelling problems concern the nonlinear stage of MI~\cite{biondini2018universal} along with the existence of rogue waves~\cite{el2016dam} in the 1D droplet environment. 
These investigations would require the use of  multiscale expansion techniques and potential reconstruction of solutions thereof in the present setting. Another intriguing problem in the context of the observed traveling DSW is to study the existence of generalized heteroclinic connections between periodic and homogeneous backgrounds observed elsewhere \cite{hoefer2019modulation,sprenger2017shock}.
Another extension of immediate interest, would be the study of oblique~\cite{hoefer2017oblique} and radial DSWs~\cite{hoefer2006dispersive} in two-dimensional droplet settings. 
Investigations, that thus far have only been conducted in setups lacking the inclusion of quantum fluctuations. 
Finally, exploring the growth of correlations and entanglement accompanying the emergence of DSWs in this attractive environment through nonperturbative approaches~\cite{mistakidis2021formation,englezos2023correlated} is certainly desirable. 


\begin{acknowledgments}
  We thank Dr. Patrick Sprenger and Prof. Mark A. Hoefer for inspiring discussions on the subject matter of Whitham shocks. S.C. thanks Missouri University of science and technology for their hospitality, where a part of this work was discussed. {This material is based upon work supported by the
U.S. National Science Foundation under the awards
PHY-2110030 and DMS-2204702 (PGK).}
\end{acknowledgments}
\label{Section-5}

\appendix
\section{Riemann invariants and Whitham modulation equations at the linear and solitonic edges}
\label{appendix:Whitham_eq}
Explicit formulae for the Riemann invariants $r_{1,2}$ [c.f. Eqs.~\eqref{Hydrodynamic-reduction}] are given below
\begin{align}
\label{Expression-Riemann}
   r_{1,2} &=\frac{u}{2}\\\nonumber &\mp \frac{1}{2}\left[\sqrt{2A}\sqrt{2A-1}-\frac{1}{2}{\rm ln}(4A-1+2\sqrt{2A}\sqrt{2A-1})\right]
\end{align}
where, $A=\sqrt{\rho}$ is the amplitude of the wave function. The Riemann invariant formulation is important to study the Whitham-El simple wave DSW theory (see \cite{el2005resolution,hoefer2014shock,el2007theory}).

Furthermore, in this section, we derive simple wave ordinary differential equations (ODE) from the Whitham equations at the linear and solitonic edges that can be used to predict the associated edge speeds. We first examine this Whitham equation system (Eqs.~\eqref{Whitham-set1},\eqref{Consn-waves}) in the vicinity of the 2-DSW linear edge. Downstream to this linear edge, the one-phase modulated zone transitions to the dispersionless (oscillation free) limit of the DH equation system Eqs.~\eqref{Quasilinear-system-DH-1}-\eqref{Quasilinear-system-DH-2}.

Here, it is well known that the equation system reduces to a system of three equations given by (see \cite{hoefer2014shock,el2005resolution} for a discussion for generalized Schrödinger-type models)
\begin{subequations}
\begin{align}
\label{wavenumber-lin-edge}
    &k_t+(\omega_0(k,\overline{\rho},\overline{k}))_x=0,\\
    \label{density-lin-edge}
    &\overline{\rho}_t+(\overline{\rho}\;\overline{u})_x=0,\\
    \label{velocity-lin-edge}
&\overline{u}_t+\overline{u}\;\overline{u}_x+f^{\prime}(\overline{\rho})\overline{\rho}_x=0.   
\end{align}
\end{subequations}
where $k$ is the wavenumber of small amplitude oscillations in the vicinity of the linear edge, $\omega_0$ is the linear dispersion function (Eq.~\eqref{Disp-relation-linear-edge}) and   $f(\overline{\rho})=\overline{\rho}-\sqrt{\overline{\rho}}$. Note that in this linear limit $\overline{G(\rho,u)}=G(\overline{\rho},\overline{u})$, for some generic function $G$ \cite{el2005resolution}. Furthermore, for the right propagating 2-DSW we observe that the positive branch of Eq.~\eqref{Disp-relation-linear-edge} describes the appropriate linear dispersion relation. 
Finally, we note that Eqs.~\eqref{density-lin-edge},\eqref{velocity-lin-edge} are decoupled from \eqref{wavenumber-lin-edge} in the linear limit, which simplifies their simultaneous solution.

We initiate the simultaneous solution of Eqs.~\eqref{wavenumber-lin-edge}-\eqref{velocity-lin-edge} by invoking the 2-DSW relation (see also \cite{hoefer2014shock,crosta2012whitham}) given by $\overline{u}=\int_{\rho_0^{(2)}}^{\overline{\rho}} \frac{c}{\rho^{\prime}} d\rho^{\prime}$,
where $c$ is the speed of sound. Substituting this averaged velocity in the vicinity of the linear edge into the linear modulation system leads to a system of two equations 
\begin{subequations}
\begin{align}
\label{k-linear-edge-2}
    &k_t+(\omega_0)_x=0,\\
   \label{rho-linear-edge-2} &\overline{\rho}_t+\mathcal{V}(\overline{\rho})\overline{\rho}_x=0,
\end{align}
\end{subequations}
where $\mathcal{V}=\int_{\rho_0^{(2)}}^{\rho} \frac{c}{\rho^{\prime}} d\rho^{\prime}+c$. The reduced hyperbolic system in Eqs.~\eqref{k-linear-edge-2}-\eqref{rho-linear-edge-2} possesses two characteristic velocities ($\partial_k\omega_0$ and $\mathcal{V}$). One can derive the integral curve of this reduced two-equation Whitham system by pre-multiplication with its left eigenvector $[\partial_k\omega_0-\mathcal{V},\;\; \partial_{\overline{\rho}}\omega_0]^T$ which yields the first simple wave ODE
\begin{equation}
\label{Integral-curve}
    \frac{dk}{d\overline{\rho}}=\frac{ck/\overline{\rho}+\partial_{\overline{\rho}}(k\sqrt{k^2/4+c^2})}{{c-\partial_k}(k\sqrt{k^2/4+c^2})}.
\end{equation}
{For convenience, one instead could look at a related simple wave ODE for the scaled phase speed variable $\alpha=(1/c)\sqrt{\frac{k^2}{4}+c^2}=\sqrt{\frac{k^2}{4c^2}+1}$, which {upon incorporating the relationship ${d\alpha}/{d\overline{\rho}}={k}/{4c^2\alpha}\bigg({dk}/{d\overline{\rho}}-k{c_{\rho}}/{c}\bigg)$} gives}
\begin{equation}
\label{alpha-rho-ODE}
   \frac{d\alpha}{d\overline{\rho}}=-\frac{(1+\alpha)}{2}\bigg(\frac{1}{\overline{\rho}}+\frac{2\alpha-1}{2\alpha+1}\frac{f_{\overline \rho\;\overline \rho}}{f_{\overline \rho}}\bigg). 
\end{equation}
The ODE above is subject to the condition that the wavenumber (scaled phase speed) at the 2-DSW soliton edge is zero (one), i.e. $k(\rho_m)=0$ ($\alpha(\rho_m)=1$) \cite{hoefer2014shock,el2005resolution}. This (\eqref{alpha-rho-ODE}) can be mapped to an appropriate initial value problem by defining the forward ``time" like variable
\begin{equation}
    \tau=\rho_m-\overline{\rho},
\end{equation}
for which $\tau(\rho_m)=0$ and $\tau(\rho_0^{(2)})=\rho_m-\rho_0^{(2)}$, and the corresponding ODE for $\alpha(\tau)$ being
\begin{equation}
    \frac{d\alpha}{d\tau}= \frac{(1+\alpha)}{2}\bigg(\frac{1}{\rho_m-\tau}+\frac{2\alpha-1}{2\alpha+1}\frac{f_{\rho\rho}(\tau)}{f_{\rho}(\tau)}\bigg),
\end{equation}
that is supplemented with initial condition $\alpha(0)=1$. Upon integrating this IVP in time $\tau$ with a RK-4 time-stepper, we obtain $\alpha(\rho_0^{(2)})$. 
Thereafter, at the linear edge, since $\overline{u}(\rho_0^{(2)})=0$, we obtain the associated wavenumber 
\begin{equation}
    \label{wavenumber-linear-edge}
    k(\rho_0^{(2)})=2(c(\rho_0^{(2)}))\sqrt{(\alpha(\rho_0^{(2)}))^2-1}.
\end{equation} Substituting this into the expression for the group velocity $\partial_k\omega_0(k(\rho_0^{(2)}),\rho_0^{(2)})$ we get the expression for the speed of the linear edge of the 2-DSW

\begin{equation}
   v_{+}(\rho_0^{(2)})= \bigg((2c_0^{(2)}(\alpha_0^{(2)})^2-c_0^{(2)})/\alpha_0^{(2)}\bigg),
\end{equation}
where $\alpha(\rho_0^{(2)})$ and $c(\rho_0^{(2)})$ are written as $\alpha_0^{(2)}$ and $c_0^{(2)}$ respectively.

Having determined both the (a) wavenumber and (b) the speed $v_{+}$ at the linear edge, we describe the calculation of the soliton edge speed $v_{-}$ of the 2-DSW. The associated task is to now determine the phase speed of the soliton edge, which would correspond to the DSW left edge speed. In this vein, it is beneficial to utilize the conjugate wavenumber ($\Tilde{k}$) approach (see \cite{el2005resolution} for details), where $\Tilde{k}$ plays a similar role to the wave amplitude. Moreover, this defines the conjugate angular frequency $\Tilde{\omega}(\Tilde{k},\overline{\rho})=-i\omega_0(i\Tilde{k},\overline{\rho})$, which translates to
\begin{align}
    \Tilde{\omega}&=\overline{u}\Tilde{k}+\Tilde{k}\sqrt{c^2-\frac{\Tilde{k}^2}{4}},\\\nonumber
    &=\overline{u}\Tilde{k}+\omega_0(\Tilde{k},\overline{\rho}),
\end{align}
The conjugate phase speed at the soliton edge of the 2-DSW then yields the soliton phase speed
\begin{equation}
    v_{-}=\frac{\Tilde{\omega}(\rho_m,\Tilde{k}_m)}{\Tilde{k}_m}=u_m+c_m\sqrt{1-\frac{\Tilde{k}_m^2}{4c_m^2}},
\end{equation}
where $c_m=c(\rho_m)$ and we have from the 2-DSW jump condition
\begin{equation}
    u_m=\int_{\rho_0^{(2)}}^{\rho_m}\frac{c(\rho^{\prime})}{\rho^{\prime}} d\rho^{\prime} 
\end{equation}
Thus, what remains to be determined is the conjugate wavenumber $\Tilde{k}_m$ at the soliton edge; the integral curve governing the ``dynamics" of in the $k=0$ plane can be derived from the second simple wave ODE
 \begin{figure}
\vspace{4mm}
\centering    \includegraphics[width=1.0\linewidth]{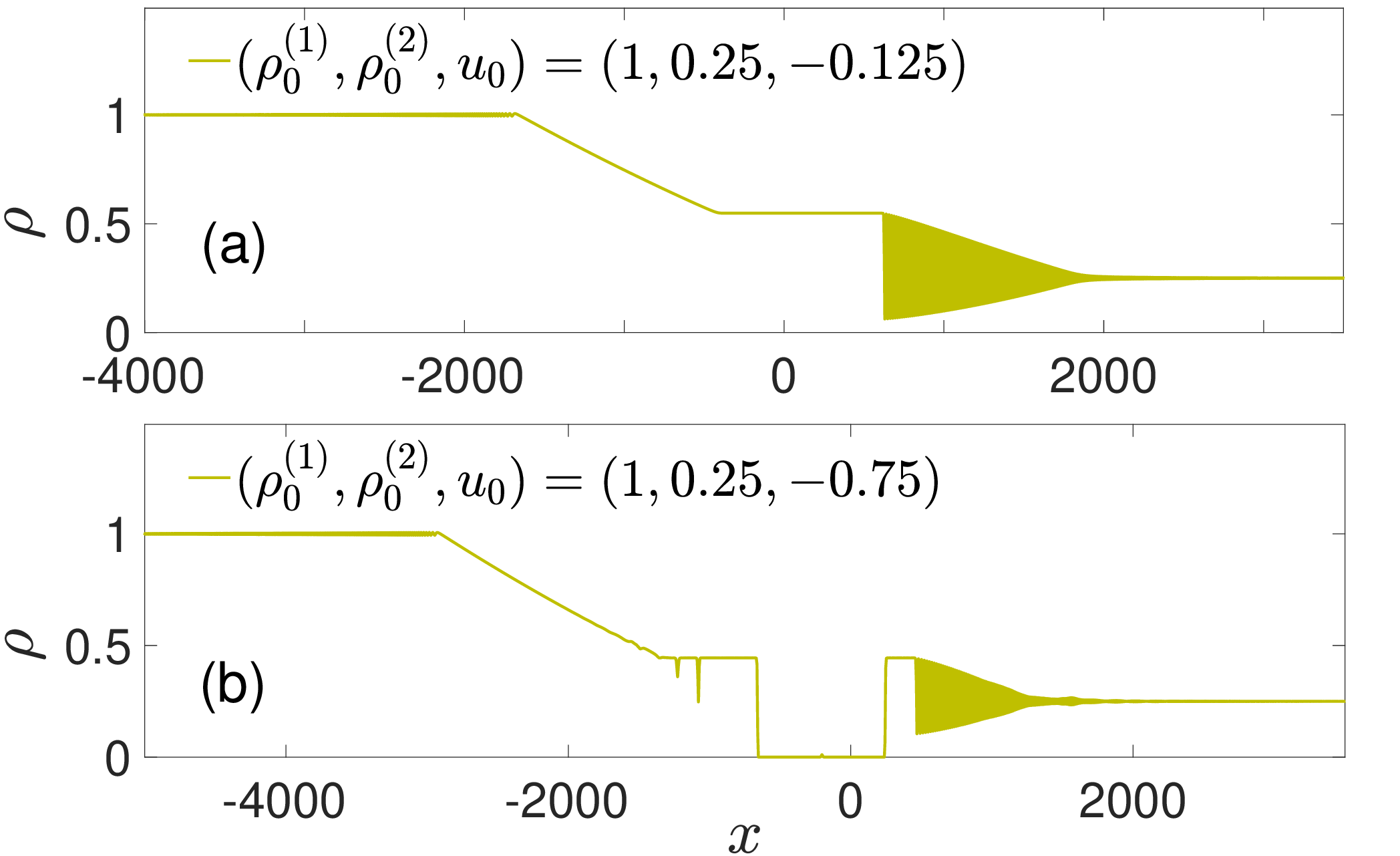}
\caption{(a), (b) Density profiles at $t=2000$ for different $(\rho_0^{(1)},\rho_0^{(2)},u_0$ (see legends).  
The emergent wave patterns having velocities below $u_c$ consist of (a) a counterpropagating rarefaction and a DSW. and (b) a counterpropagating kink-antikink pair and a DSW. }
\label{fig:9}
\end{figure}
\begin{equation}
\label{Conjugate-wavenumber-ODE-relation}
    \frac{d\Tilde{k}}{d\overline{\rho}}=\frac{c(\overline{\rho})\Tilde{k}/\overline{\rho}+\partial_{\overline{\rho}}(\Tilde{k}\sqrt{c^2-\frac{\Tilde{k}^2}{4}})}{c(\overline{\rho})-\partial_{\Tilde{k}}(\Tilde{k}\sqrt{c^2-\frac{\Tilde{k}^2}{4}})},
\end{equation}
and \eqref{Conjugate-wavenumber-ODE-relation} is subject to initial condition $\Tilde{k}(\rho_0^{(2)})=0$ (reflects the zero amplitude condition at linear edge). Note that  one typically solves the IVP for the associated conjugate, scaled phase speed $\Tilde{\alpha}=\sqrt{1-\frac{\Tilde{k}^2}{4c^2}}$. The associated ODE for the conjugate phase speed can be derived to have the same form as Eq.~\eqref{alpha-rho-ODE}, and is supplemented with ``initial condition" $\Tilde{\alpha}(\rho_0^{(2)})=1$ . This conjugate ODE is integrated forward in time with an RK-4 time stepper to obtain $\Tilde{\alpha}(\rho_m)=\Tilde{\alpha}_m$. Ultimately, we can then obtain the soliton edge speed as $v_{-}=u_m+c_m\Tilde{\alpha}_m$. 

 \section{Riemann problems with non-trivial velocity $u_0^{(1)}\neq 0$}
 
We are interested in studying wave pattern formation under the active competition of mean-field and LHY contributions of Eq.~\eqref{Riemann-step-GP} as in the main text but for finite velocity $u_0<0$. 
Recall that in classical dispersive Eulerian hydrodynamics~\cite{el1995decay}, $u_0<0$ implies that the generated wave patterns get closer to the vacuum state, and, for $|u_0|\gg1$ speeds, one expects the generation of  counterpropagating rarefaction waves on the intermediate vacuum state $\rho_m=0$. {However, a far richer phenomenology occurs for the model under consideration in the LHY dominated regime. In what follows, we explicate this phenomenology 
 with two representative examples. To set the stage, we decrease $u_0<0$ starting from the dam break limit ($u_0=0$). The progressive reduction of $u_0$ initially leads to a corresponding decrease of the intermediate density $\rho_m$ (predicted by the Whitham-El closure method) between a rarefaction wave and a DSW (c.f. Fig.~\ref{fig:9}(a)). It is possible to predict the critical velocity $u_0^{(1)}=u_c$ for which $\rho_m= 0.25$, corresponding to the hyperbolic threshold.  For instance, in the case of $(\rho_0^{(1)},\rho_0^{(2)})=(1,0.25)$, this critical threshold as predicted by the closure method yields $u_c= \frac{1}{2}{\rm ln}(3+2\sqrt{2})-\sqrt{2}\approx -0.53$, and our eGPE simulations agree with this prediction to about $5\%$. Below $u_0^{(1)}<u_c$, a far richer wave pattern emerges due to the generation of counterpropagating kink-antikink pairs. These are connected to the external flow through a rarefaction and a 2-DSW respectively, as depicted in Fig.~\ref{fig:9}(b). Such patterns, which have also previously arisen
 in studies of the eGPE~\cite{katsimiga2023interactions},
 merit further investigation in their own right that is deferred
 to future studies.}

\bibliography{Sept_16/SHOCK_WAVES_QD_REVIEW}
\end{document}